\documentclass[preprint,12pt]{elsarticle}

\usepackage{amssymb}
\usepackage{amsmath}
\usepackage{mathrsfs} 
\usepackage[version=3]{mhchem} % Formula subscripts using \ce{}
\usepackage{graphicx} %
\usepackage{float} %
\usepackage{subfigure} %
\usepackage{multirow}
\usepackage{multicol}
\usepackage{booktabs}
\usepackage[utf8]{inputenc}
\usepackage{diagbox}
\usepackage{afterpage}
\usepackage{caption}
\usepackage{subcaption}

\newcommand*{\citen}[1]{%
  \begingroup
    \romannumeral-`\x % remove space at the beginning of \setcitestyle
    \setcitestyle{numbers}%
    \cite{#1}%
  \endgroup   
}
\journal{Computer Physics Communications}

\begin{document}

\begin{frontmatter}



\title{Functional Analytic Derivation and CP2K Implementation of the SCCS Model Based on the Solvent-Aware Interface}

 \author{Ziwei Chai\corref{cor1}\fnref{label1}}
 \ead{ziwei.chai@chem.uzh.ch}
 \author{Sandra Luber\fnref{label1}}
 \cortext[cor1]{Corresponding author.}
 \affiliation[label1]{organization={Department of Chemistry, University of Zurich},%Department and Organization
            addressline={Winterthurerstrasse 190}, 
            city={Zurich},
            postcode={8057}, 
            state={Zurich},
            country={Switzerland}}

%% Abstract
\begin{abstract}
In the self-consistent continuum solvation (SCCS) approach (\textit{J. Chem. Phys.} 136, 064102 (2012)), the analytical expressions of the local solute-solvent interface functions determine the interface function and dielectric function values at a given real space position based solely on the electron density at that position, completely disregarding the surrounding electron density distribution. Therefore, the low electron density areas inside the solute will be identified by the algorithm as regions where implicit solvent exists, resulting in the emergence of non-physical implicit solvent regions within the solute and even potentially leading to the divergence catastrophe of Kohn-Sham SCF calculations. We present a new and efficient SCCS implementation based on the solvent-aware interface (\textit{J. Chem. Theory Comput.} 15, 3, 1996–2009 (2019)) which addresses this issue by utilizing a solute-solvent interface function based on convolution of electron density in the CP2K software package, which is based on the mixed Gaussian and plane waves (GPW) approach. Starting with the foundational formulas of SCCS, we have rigorously and meticulously derived the contributions of the newly defined electrostatic energy to the Kohn-Sham potential and the analytical forces. This comprehensive derivation utilizes the updated versions of the solute-solvent interface function and the dielectric function, tailored to align with the specifics of the GPW implementation. %The accuracy of the derivation and implementation of the analytical forces has been verified by comparing the analytical forces and numerical differential forces.
Our implementation has been tested to successfully eliminate non-physical implicit solvent regions within the solute and achieve good SCF convergence, as demonstrated by test results for both bulk and surface models, namely liquid H\(_2\)O, titanium dioxide, and platinum.
\end{abstract}

\begin{graphicalabstract}
\includegraphics{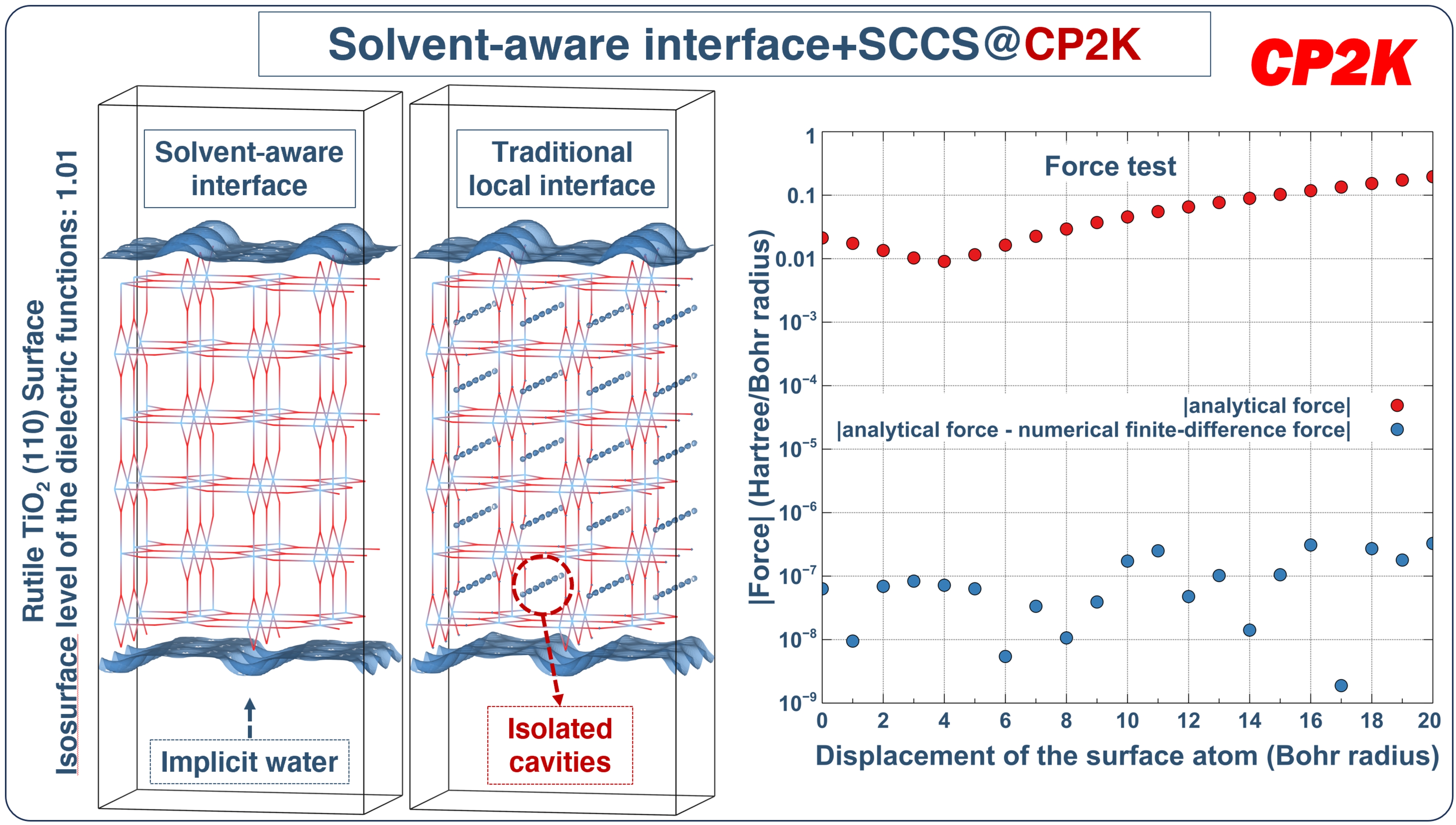}
\end{graphicalabstract}

%% Keywords

\end{frontmatter}

\section{Introduction}
\label{sec1}
Solvents play a critical role in chemical, biological, and industrial processes\cite{noauthor_introduction_2002,orozco_theoretical_2000,ringe_implicit_2022}. In the computational simulation of the processes, solvents are commonly modeled as explicit atomistic models, fully implicit dielectric continuum, or coarse-grained models between the two\cite{car_unified_1985,warshel_theoretical_1976,frenkel_understanding_2023,allen_computer_2017,gros_ab_2022,ringe_implicit_2022}.\\
When studying the effects of solvents on chemical reactions\cite{sandra2017_jctc1,sandra_jctc2_2017}, catalysis\cite{mauro_jctc,mauro_acscata,fabrizio_110_explicit_solvent,fabrizio_ruo2_explicit_solvent}, and excited-state\cite{momir_2,momir_1,johann_1,lukas_1,eva_jctc_2024}, etc., processes, explicit solvent models are often used. However, to calculate the system's macroscopic thermodynamic quantities, such as free energies, which can directly be compared to experimental data, from a vast space of micro configurations of explicit atomic models, one needs to perform extensive sampling of them which is computationally expensive. In addition, the quantum mechanical description of such models by using quantum chemical methods such as density functional theory (DFT) will exacerbate the situation, as even by using modern advanced supercomputers and acceleration algorithms they are still computationally very demanding for large systems\cite{ringe_implicit_2022,bruix_first-principles-based_2019}.\\
From the explicit representation of solvent molecules, the degrees of freedom of nuclei of solvent molecules can be removed partly as in RISM (the reference interaction site model) -type models or even removed further to arrive at fully implicit models of the solvents\cite{ringe_implicit_2022}. Among the fully implicit models, continuum solvation models are popular. In these models, the solvent is represented as a polarizable continuum surrounding the solute (e.g., a solid slab + adsorbed molecules or ions), depicted by the dielectric function $\epsilon(\boldsymbol{r})$\cite{tomasi_molecular_1994,ringe_implicit_2022,fattebert_density_2002,scherlis_unified_2006,andreussi_revised_2012,yin_periodic_2017}. Fattebert and Gygi formulated the dielectric function as a local function of the electron density\cite{fattebert_density_2002}, and Andreussi et al. proposed an improved form of the local function\cite{andreussi_revised_2012}. The electrostatic potential can then be calculated by solving the generalized Poisson-Boltzmann equation for the total solute charge density embedded in a given spatial distribution of the relative permittivity. The statistical average (mean field) electrostatic response of the solvent to a given electronic + ionic charge distribution for a solute geometry is coded in the resulting total electrostatic potential. In practice, the generalized Poisson-Boltzmann equation is usually numerically solved. In the method known as the self-consistent continuum solvation (SCCS) model developed by Andreussi et al.\cite{andreussi_revised_2012}, the generalized Poisson equation is reformulated to resemble a vacuum-like Poisson equation. This reformulation allows for the equation to be solved through an iterative process. Non-electrostatic energy corrections concerning solute cavity formation, exchange repulsion, dispersion interactions, and thermal motion of the solute can also be considered\cite{scherlis_unified_2006,cococcioni_electronic-enthalpy_2005,andreussi_revised_2012,ringe_implicit_2022}.\\
According to Refs. \citen{andreussi_revised_2012} and \citen{fattebert_density_2002}, the dielectric function value at position \(\boldsymbol{r}\) ($\epsilon\left(\rho^{e l}(\boldsymbol{r})\right)$) depends only on the electron density at that position, not on densities at other positions. Therefore, if the electron density at \(\boldsymbol{r}\) in the solute (\(\rho^{el}(\boldsymbol{r})\)) falls between the threshold \(\rho^{max}\) and \(\rho^{min}\) or below \(\rho^{min}\) which delineate the boundary between the pure solute and the pure implicit solvent during SCF wave function optimization, that position will be identified as being in the solute-solvent interface region or even in the pure solvent region, respectively. Unphysical regions, where the dielectric function values exceed 1 in the solute body indicating the existence of the implicit solvent, often appear as pockets \cite{fattebert_density_2002} or isolated cavities, based on our experience. When the electron density threshold parameters that determine the solute-solvent boundary are set high or the calculations involve the grand canonical SCF approach\cite{sundararaman_grand_2017}, the aforementioned unphysical implicit solvent regions in the solute body are more likely to appear, in our experience. This can potentially lead to divergences in the iterations of the generalized Poisson equation and (grand canonical) Kohn-Sham SCF\cite{sundararaman_grand_2017}. The solvent-aware algorithm proposed by Andreussi et al.\cite{fattebert_density_2002} was designed to solve this problem. In this algorithm, the dielectric function at \(\boldsymbol{r}\) depends not only on the electron density at \(\boldsymbol{r}\) but also on the electron densities in the vicinity of this position. Non-locality is introduced by calculating the convolution of the monotonically decreasing complementary error function (often referred to as 'erfc') with the interface function \(s(\boldsymbol{r})\), where \(s(\boldsymbol{r}) = 0\) ($\boldsymbol{r}$ is in the pure implicit solvent) if \(\rho^{el}(\boldsymbol{r}) < \rho^{min}\), \(s(\boldsymbol{r}) = 1\) ($\boldsymbol{r}$ is in the pure solute) if \(\rho^{el}(\boldsymbol{r}) > \rho^{max}\), and \(s(\boldsymbol{r}) \in (0,1)\) ($\boldsymbol{r}$ is in the transition region) if \(\rho^{min} \leq \rho^{el}(\boldsymbol{r}) \leq \rho^{max}\). The scaled difference between a threshold parameter and the convolution value serves as the input for the error function (often referred to as 'erf'), determining the mixing coefficient of the original interface function \(s(\boldsymbol{r})\) with 1. This method can successfully identify and eliminate unphysical pockets and isolated cavities with relative permittivity greater than 1 inside solutes. The corresponding electrostatic energy and analytical forms for its contributions to the Kohn-Sham potentials and atomic forces can be derived and calculated \cite{andreussi_solvent-aware_2019}.\\
In this paper, we will first discuss and present the rigorous and meticulous theoretical derivations of the modified SCCS approach based on the solvent-aware solute-solvent interface, starting with the foundational formulas of SCCS from scratch. We have implemented this approach in the CP2K software package which can efficiently perform atomistic simulations of various systems in parallel using a combination of multi-threading, MPI, and CUDA. We then present a comparison between calculated analytical forces and numerical forces. At the end, we present the test results for the new implementation of the modified SCCS approach in predicting solvent-solute interfaces across a series of different types of systems and under various computational settings, as well as the improvement in SCF convergence.\\
The paper is organized as follows: In Section \ref{sec2} we present the theory of the solvent-aware interface and the modified SCCS compatible with the GPW approach used in the CP2K software package. In Section \ref{subsec2.1} and Section \ref{subsec2.2} about the solute-solvent interface approaches, Andreussi et al.’s local solute-solvent interface and the non-local interface in the solvent-aware algorithm, respectively are presented. In Section \ref{subsec2.3}, the theory of the modified SCCS with solvent-aware interfaces including the generalized Poisson equation (Section \ref{subsec2.3.1}), electrostatic energy (Section \ref{subsec2.3.2}), and contribution to Kohn-Sham potentials (Section \ref{subsec2.3.3}), contribution to analytical forces (Section \ref{subsec2.3.4}) is discussed. In Section \ref{sec3} we outline the computational settings and parameters which were adopted in the tests. In Section \ref{sec4} we present the test results including the analytical force validation (Section \ref{subsec4.1}), the test results for solute-solvent boundary and SCF convergence (Section \ref{subsec4.2}). In Section \ref{sec5} we provide the conclusions.\\
\section{Theory}
\label{sec2}
%%%%%%%%%%%%%%%%%%%%%%%%%%%%%%%%%%%%%%%%%%%%%
\renewcommand{\thefootnote}{\fnsymbol{footnote}}
\subsection{Andreussi et al.'s local solute-solvent interface}
\label{subsec2.1}
In Andreussi et al.'s local solute-solvent interface\cite{andreussi_revised_2012}, the value of the dielectric function, denoted as $\epsilon$, at a given position in real space $\boldsymbol{r}$, is calculated based on the value of the interface function $s$ at $\boldsymbol{r}$ and the dielectric constant of the bulk solvent $\epsilon_0$:
\begin{equation}\label{eqs1}
\epsilon(\boldsymbol{r})=\epsilon\left(\rho^{e l}(\boldsymbol{r})\right)=e^{(\ln \epsilon_0)\left(1-s\left(\rho^{e l}(\boldsymbol{r})\right)\right)}.
\end{equation}
From Eq. \eqref{eqs1}, we see that at each position $\boldsymbol{r}$, narrowly speaking, $\epsilon$ is a function of $s$ instead of a non-local functional of the $s(\boldsymbol{r^{\prime}})$ function. ($\boldsymbol{r^{\prime}}$ also runs over all points in space and is used to differentiate it from $\boldsymbol{r}$.) The interface function $s$ at $\boldsymbol{r}$ is a local function of $T(\boldsymbol{r})$:
\begin{equation}\label{eqs2}
s(\boldsymbol{r})=s_{\rho^{max }, \rho^{min }}\left(\rho^{e l}(\boldsymbol{r})\right)=\left\{\begin{array}{cl}
0 & \text{if } \rho^{e l}(\boldsymbol{r})<\rho^{min } \\
1-T\left(\ln \rho^{e l}(\boldsymbol{r})\right) & \text{if } \rho^{min } \leq \rho^{e l}(\boldsymbol{r}) \leq \rho^{max },\\
1 & \text{if } \rho^{e l}(\boldsymbol{r})>\rho^{max }
\end{array}\right.
\end{equation}
where $\rho^{max }$ and $\rho^{min }$ are maximum and minimum thresholds of the electron density $\rho^{e l}(\boldsymbol{r})$ to define the spatial boundaries of the pure solute and the pure implicit solvent, respectively. The function $T$ at $\boldsymbol{r}$ is locally determined by $\rho^{e l}(\boldsymbol{r})$ according to the form of the $T(x)$ function:
\begin{equation}\label{eqs3}
T(x)=\frac{\ln \rho^{max }-x}{\ln \rho^{max }-\ln \rho^{min }}-\frac{1}{2 \pi} \sin \left(2 \pi \frac{\ln \rho^{max }-x}{\ln \rho^{max }-\ln \rho^{min }}\right).
\end{equation}
From Eq. \eqref{eqs1} to Eq. \eqref{eqs3}, it is clear that both the interface function $s$ and the dielectric function $\epsilon$ at position $\boldsymbol{r}$ are locally determined by $\rho^{e l}(\boldsymbol{r})$. By plotting $T(\ln \rho^{e l}(\boldsymbol{r}))$, it becomes evident that the function $T(\ln \rho^{e l}(\boldsymbol{r}))$ smoothly decreases from 1 to 0 as $\ln \rho^{e l}(\boldsymbol{r})$ increases from $\ln \rho^{min}$ to $\ln \rho^{max}$. As a result, the interface function $s(\boldsymbol{r})$ smoothly increases from 0 to 1 as $\rho^{el}$ increases from $\rho^{min}$ to $\rho^{max}$, and the dielectric function $\epsilon(\boldsymbol{r})$ decreased from $\epsilon_0$ to 1 as $\rho^{el}$ increases from $\rho^{min}$ to $\rho^{max}$.
From Eqs.~\ref{eqs1}--\ref{eqs3}, which clearly establish the dependence of the dielectric function $\epsilon$ at position $\boldsymbol{r}$ on the electron density $\rho^{e l}(\boldsymbol{r})$, the first derivative of the dielectric function with respect to electron density can be easily calculated as follows, when $\rho^{min} \leq \rho^{e l}(\boldsymbol{r}) \leq \rho^{max}$:
\begin{equation}\label{eqs4}
\begin{aligned}
& \frac{d \epsilon\left(\rho^{el}\right)}{d \rho^{el}}(\boldsymbol{r}) \\
& = \epsilon(\boldsymbol{r}) (\ln \epsilon_0)(-1) \frac{d s\left(\rho^{el}\right)}{d \rho^{el}}(\boldsymbol{r}) \\
& = \epsilon(\boldsymbol{r}) (\ln \epsilon_0)(-1)(-1) \frac{d T}{d \rho^{el}}(\boldsymbol{r}) \\
& = \epsilon(\boldsymbol{r}) \left(\ln \epsilon_0\right) \left( \frac{-1}{\ln \rho^{max} - \ln \rho^{min}} \frac{d \ln \rho^{el}}{d \rho^{el}} \right. \\
& \quad + \left. \cos \left( 2 \pi \frac{\ln \rho^{max} - \ln \rho^{el}}{\ln \rho^{max} - \ln \rho^{min}} \right) \frac{1}{\ln \rho^{max} - \ln \rho^{min}} \frac{d \ln \rho^{el}}{d \rho^{el}} \right) \\
& = \epsilon(\boldsymbol{r}) \frac{1}{\rho^{el}} \left(\ln \epsilon_0\right) \frac{1}{\ln \rho^{max} - \ln \rho^{min}} \\
& \quad \left( \cos \left( 2 \pi \frac{\ln \rho^{max} - \ln \rho^{el}}{\ln \rho^{max} - \ln \rho^{min}} \right) - 1 \right).
\end{aligned}
\end{equation}
When $\rho^{e l}(\boldsymbol{r}) < \rho^{min}$ or $\rho^{e l}(\boldsymbol{r}) > \rho^{max}$, the dielectric function $\epsilon(\boldsymbol{r})$ remains constant, and its derivative $\frac{d \epsilon\left(\rho^{e l}\right)}{d \rho^{e l}}(\boldsymbol{r})$ equals 0. Referring to Eq.~\ref{eqs4}, this derivative also equals 0 at $\rho^{e l}(\boldsymbol{r}) = \rho^{min}$ or $\rho^{e l}(\boldsymbol{r}) = \rho^{max}$. Therefore, $\frac{d\epsilon}{d\rho^{e l}}(\boldsymbol{r})$ is a continuous function at each position $\boldsymbol{r}$.\\
In mathematics, a functional refers to the mapping between a function and a scalar. To make derivations clearer and more understandable, we will henceforth use $i(\boldsymbol{r})_{[j(\boldsymbol{r}^{\prime})]}$ to denote that the value of $i$ at $\boldsymbol{r}$ is a functional of the function $j(\boldsymbol{r}^{\prime})$. Similarly, we will use $i(\boldsymbol{r})_{[i(j(\boldsymbol{r}^{\prime}))]}$ to indicate that the value of $i$ at $\boldsymbol{r}$ is formally a functional of $i(\boldsymbol{r}^{\prime})$, where $i(\boldsymbol{r}^{\prime})$ is locally dependent on $j(\boldsymbol{r}^{\prime})$. We use $\frac{\delta i(\boldsymbol{r})_{[j(\boldsymbol{r}^{\prime})]}}{\delta j(\boldsymbol{r}^{\prime})}$ or $\frac{\delta i(\boldsymbol{r})_{[i(j(\boldsymbol{r}^{\prime}))]}}{\delta j(\boldsymbol{r}^{\prime})}$ to denote the functional derivative of $i(\boldsymbol{r})$ with respect to the function $j(\boldsymbol{r}^{\prime})$. Since $s(\boldsymbol{r})$ and $\epsilon(\boldsymbol{r})$ only depend on $\rho^{e l}$ at $\boldsymbol{r}$, the functional derivatives $\frac{\delta s(\boldsymbol{r})_{[s(\rho^{e l}(\boldsymbol{r}^{\prime}))]}}{\delta \rho^{e l}(\boldsymbol{r}^{\prime})}$ and $\frac{\delta \epsilon(\boldsymbol{r})_{[\epsilon(\rho^{e l}(\boldsymbol{r}^{\prime}))]}}{\delta \rho^{e l}(\boldsymbol{r}^{\prime})}$ can be calculated as follows:
\begin{equation}\label{eqs5}
\frac{\delta s(\boldsymbol{r})_{\left[s\left(\rho^{e l}\left(\boldsymbol{r}^{\prime}\right)\right)\right]}}{\delta \rho^{e l}\left(\boldsymbol{r}^{\prime}\right)} = \frac{\delta s(\boldsymbol{r})_{\left[s\left(\rho^{e l}\left(\boldsymbol{r}^{\prime}\right)\right)\right]}}{\delta s\left(\boldsymbol{r}^{\prime}\right)} \frac{d s\left(\rho^{e l}\right)}{d \rho^{e l}}\left(\boldsymbol{r}^{\prime}\right) = \delta\left(\boldsymbol{r}-\boldsymbol{r}^{\prime}\right) \frac{d s\left(\rho^{e l}\right)}{d \rho^{e l}}\left(\boldsymbol{r}^{\prime}\right),
\end{equation}
and
\begin{equation}\label{eqs6}
\frac{\delta \epsilon(\boldsymbol{r})_{[\epsilon(\rho^{e l}(\boldsymbol{r}^{\prime}))]}}{\delta \rho^{e l}(\boldsymbol{r}^{\prime})} = \delta(\boldsymbol{r}-\boldsymbol{r}^{\prime}) \frac{d \epsilon(\rho^{e l})}{d \rho^{e l}}(\boldsymbol{r}^{\prime}),
\end{equation}
according to the chain rule for functional derivatives (detailed in Section \ref{app7}). From Eq.~\ref{eqs1}, $\frac{\delta s(\boldsymbol{r})_{[s(\rho^{e l}(\boldsymbol{r}^{\prime}))]}}{\delta \rho^{e l}(\boldsymbol{r}^{\prime})}$ can be expressed as:
\begin{equation}\label{eqs7}
\frac{\delta s(\boldsymbol{r})_{[s(\rho^{e l}(\boldsymbol{r}^{\prime}))]}}{\delta \rho^{e l}(\boldsymbol{r}^{\prime})} = \frac{\delta \epsilon(\boldsymbol{r})_{[\epsilon(\rho^{e l}(\boldsymbol{r}^{\prime}))]}}{\delta \rho^{e l}(\boldsymbol{r}^{\prime})} \frac{1}{\epsilon(\boldsymbol{r}) \ln \epsilon_0(-1)}.
\end{equation}
\subsection{The non-local solute-solvent interface in the solvent-aware algorithm}
\label{subsec2.2}
At each position $\boldsymbol{r}$, the solvent-aware algorithm mixes 1 and the local interface function $s(\boldsymbol{r})$ in a certain ratio to get the new interface function $\hat{s}(\boldsymbol{r})$\cite{andreussi_solvent-aware_2019}
\begin{equation}\label{eqs8}
\hat{s}(\boldsymbol{r})=s(\boldsymbol{r})+(1-s(\boldsymbol{r})) t(\boldsymbol{r})=t(\boldsymbol{r})+(1-t(\boldsymbol{r})) s(\boldsymbol{r}).
\end{equation}
$t(\boldsymbol{r})$ which was defined as\cite{andreussi_solvent-aware_2019}
\begin{equation}\label{eqs9}
t(\boldsymbol{r})=\frac{1}{2}\left[1+\operatorname{erf}\left(\frac{f(\boldsymbol{r})-f_0}{\Delta_\eta}\right)\right],
\end{equation}
and in the range of $(0,1)$ is the proportion of \(s(\boldsymbol{r}) = 1\) in $\hat{s}(\boldsymbol{r})$. $f_0$ is the threshold parameter for $f(\boldsymbol{r})$ and $\Delta_\eta$ controls the softness of function $t(\boldsymbol{r})$. According to Eq.~\ref{eqs1}, \(s(\boldsymbol{r}) = 1\) gives a value of the dielectric function at $\boldsymbol{r}$ equal to 1, which indicates that position $\boldsymbol{r}$ is within the body of the solute and is not filled with any implicit solvent. The error function \(\operatorname{erf}(x)\) increases monotonically from \(-1\) as \(x\) approaches \(-\infty\) to \(+1\) as \(x\) approaches \(+\infty\). Consequently, one can deduce that \(t(\boldsymbol{r})\) increases monotonically with respect to \(f(\boldsymbol{r})\).\\
$f(\boldsymbol{r})$ was defined as the convolution between $s$ and $u$ \cite{andreussi_solvent-aware_2019}\footnote[2]{To efficiently compute this term within the program, reliance is placed on the fast Fourier transform, as elaborated in Section \ref{subsec3.3}}:
\begin{equation}\label{eqs10}
f(\boldsymbol{r})=\int s\left(\boldsymbol{r}^{\prime}\right) u\left(\boldsymbol{r}-\boldsymbol{r}^{\prime}\right) d \boldsymbol{r}^{\prime}=s * u(\boldsymbol{r}),
\end{equation}
where the function $u$ is given by:
\begin{equation}\label{eqs11}
u(\boldsymbol{r})=\frac{1}{2 N_u} \operatorname{erfc}\left(\frac{|\boldsymbol{r}|-\alpha_\zeta R_{solv}}{\Delta_\zeta}\right).
\end{equation}
In this function, $R_{solv}$ has the physical meaning of the radius of a solvent molecule, $\Delta_\zeta$ controls the sharpness of $u(\boldsymbol{r})$, $\alpha_\zeta$ is the scaling parameter for $R_{solv}$, and $N_u$ is the normalization factor ensuring that $\int u\left(\boldsymbol{r}-\boldsymbol{r}^{\prime}\right) d \boldsymbol{r}^{\prime}=1$. As $|\boldsymbol{r}|$ increases from 0 to \(+\infty\), the complementary error function $\operatorname{erfc}\left(\frac{|\boldsymbol{r}|-\alpha_\zeta R_{solv}}{\Delta_\zeta}\right)$ decreases towards 0. Thus, $f(\boldsymbol{r})$ represents a spatial average of $s(\boldsymbol{r}')$, weighted by the proximity of $\boldsymbol{r}'$ to $\boldsymbol{r}$, where the closer $\boldsymbol{r}'$ is to $\boldsymbol{r}$, the higher the weight. From Eq. \ref{eqs8} and Eq. \ref{eqs9}, it is evident that the value of the new interface function $\hat{s}(\boldsymbol{r})$ is continuously modulated between 1 and $s\left(\boldsymbol{r}\right)$ by the weighted average $f(\boldsymbol{r})$ through $t(\boldsymbol{r})$, rather than merely depending on the electron density at position $\boldsymbol{r}$. The dielectric function is then calculated from the new interface function $\hat{s}(\boldsymbol{r})$ using Eq. \ref{eqs1}:
\begin{equation}\label{eqs12}
\hat{\epsilon}(\boldsymbol{r})=e^{\ln \epsilon_0[1-\hat{s}(\boldsymbol{r})]}.
\end{equation}
From Eq.~\ref{eqs10} and the definition of a functional derivative, one can calculate the functional derivative of \( f(\boldsymbol{r}) \) with respect to \( s(\boldsymbol{r}') \):
\begin{equation}\label{eqs13}
\frac{\delta f(\boldsymbol{r})_{\left[s\left(\boldsymbol{r}^{\prime}\right)\right]}}{\delta s\left(\boldsymbol{r}^{\prime}\right)}=u\left(\boldsymbol{r}-\boldsymbol{r}^{\prime}\right).
\end{equation}
According to Eq.~\ref{eqs9}, the derivative \( \frac{d t(f)}{d f}(\boldsymbol{r}) \) is given by:
\begin{equation}\label{eqs14}
\frac{d t(f)}{d f}(\boldsymbol{r})=\frac{1}{\sqrt{\pi}} e^{-\frac{\left(f(\boldsymbol{r})-f_0\right)^2}{\Delta_\eta{ }^2}}.
\end{equation}
Following a similar logic as in the derivation of Eq.~\ref{eqs5}, from Eq.~\ref{eqs9} one can derive:
\begin{equation}\label{eqs15}
\frac{\delta t(f(\boldsymbol{r}))_{\left[f\left(\boldsymbol{r}^{\prime}\right)\right]}}{\delta f\left(\boldsymbol{r}^{\prime}\right)}=\delta\left(\boldsymbol{r}-\boldsymbol{r}^{\prime}\right) \frac{d t(f)}{d f}\left(\boldsymbol{r}^{\prime}\right).
\end{equation}
\subsection{SCCS with solvent-aware interfaces}
\label{subsec2.3}
\subsubsection{Solving the generalized Poisson equation}
\label{subsec2.3.1}
The central idea of the solvent-aware algorithm \cite{andreussi_solvent-aware_2019} is to replace the purely local dependency of the dielectric function \(\epsilon(\boldsymbol{r})\) at position \(\boldsymbol{r}\) on the electron density $\rho^{e l}(\boldsymbol{r})$ (Eqs.~\ref{eqs1}--\ref{eqs3})—as proposed by Andreussi et al. \cite{andreussi_revised_2012}—with a non-local dependency $\hat{\epsilon}(\boldsymbol{r})_{[\rho^{e l}(\boldsymbol{r}^{\prime})]}$ (Eqs.~\ref{eqs12}, \ref{eqs8}--\ref{eqs10}, and \ref{eqs2}--\ref{eqs3}). This approach allows the algorithm to correctly identify and eliminate small isolated cavities or pockets of low electron density within a solute body, by considering electron density values nearby \(\boldsymbol{r}\) that are typically significant. This prevents these regions from being mistakenly identified as pure solvent regions or transition regions between solvent and solute. With this treatment, the corrected \(\hat{\epsilon}(\boldsymbol{r})\) can then be used to solve the generalized Poisson equation in the SCCS model\cite{fattebert_density_2002,Fattebert_2003,andreussi_revised_2012}
\begin{equation}\label{eq20}
\nabla \cdot\left(\hat{\epsilon}(\boldsymbol{r}) \nabla \phi^{tot}(\boldsymbol{r})\right)=-4 \pi \rho^{solute}(\boldsymbol{r}).
\end{equation}
It is worth noting that if a counter charge density (e.g., from a planar counter charge \cite{nattino_continuum_2018, lozovoi_reconstruction_2003}) is added to the total charge density, \( \rho^{solute}(\boldsymbol{r}) \) also includes its contribution. The electrostatic potential, $\phi^{tot}(\boldsymbol{r})$, is determined by Eq.~\ref{eq20}, which is reformulated in the SCCS approach as follows:
\begin{equation}\label{eq21}
\nabla \cdot\left(\nabla \phi^{tot}(\boldsymbol{r})\right)=\nabla^2 \phi^{tot}(\boldsymbol{r})=-4 \pi\left(\rho^{solute}(\boldsymbol{r})+\rho^{pol}(\boldsymbol{r})\right),
\end{equation}
where the polarization charge density $\rho^{p o l}(\boldsymbol{r})$ is:
\begin{equation}\label{eq22}
\rho^{pol}(\boldsymbol{r})=\nabla \cdot\left(\frac{\hat{\epsilon}(\boldsymbol{r})-1}{4 \pi} \nabla \phi^{tot}(\boldsymbol{r})\right).
\end{equation}
From Eq. \ref{eq21}--\ref{eq22}, $\rho^{p o l}(\boldsymbol{r})$ can be further reformulated as\cite{andreussi_revised_2012}:
\begin{equation}\label{eq23}
\rho^{pol}(\boldsymbol{r})=\frac{1}{4 \pi}(\nabla \ln \hat{\epsilon}(\boldsymbol{r})) \cdot\left(\nabla \phi^{tot}(\boldsymbol{r})\right)-\frac{\hat{\epsilon}(\boldsymbol{r})-1}{\hat{\epsilon}(\boldsymbol{r})} \rho^{solute}(\boldsymbol{r}) .
\end{equation}
From Eqs.~\ref{eq21} and \ref{eq23}, it is evident that \( \phi^{tot}(\boldsymbol{r}) \) is the electrostatic potential generated by the charge density \( \rho^{solute}(\boldsymbol{r}) + \rho^{pol}(\boldsymbol{r}) \), and calculating the charge density \( \rho^{pol}(\boldsymbol{r}) \) requires the electrostatic potential \( \phi^{tot}(\boldsymbol{r}) \). The self-consistent loop for solving Eq.~\ref{eq21} in the normal SCCS is mostly equivalent to the one in the solvent-aware algorithm. The only difference in this regard is whether the original or the modified dielectric function is chosen.\\
\subsubsection{Electrostatic energy}
\label{subsec2.3.2}
According to the original paper in which the normal SCCS was proposed \cite{andreussi_revised_2012}, when switching the solute's environment from vacuum to an implicit solvent, the modification for the total energy primarily involves substituting the Hartree energy of the solute in vacuum with its counterpart in the solvent. The Hartree energy of the solute in a vacuum takes the form
\begin{equation}\label{eq24}
E_{Hartree}^{old}=\frac{1}{2} \int \rho^{solute}(\boldsymbol{r}) \phi^{solute}(\boldsymbol{r}) d \boldsymbol{r},
\end{equation}
where $\phi^{solute}(\boldsymbol{r})$ is the solution to the Poisson equation for the solute in vacuum:
\begin{equation}\label{eq25}
\nabla^2 \phi^{solute}(\boldsymbol{r})=-4 \pi \rho^{solute}(\boldsymbol{r}).
\end{equation}
Meanwhile, the Hartree energy of the solute embedded in an implicit solvent environment is given by
\begin{equation}\label{eq26}
E_{Hartree}^{new}=\frac{1}{8 \pi} \int \hat{\epsilon}(\boldsymbol{r})\left|\nabla \phi^{tot}(\boldsymbol{r})\right|^2 d \boldsymbol{r}=\frac{1}{2} \int \rho^{solute}(\boldsymbol{r}) \phi^{tot}(\boldsymbol{r}) d \boldsymbol{r}.
\end{equation}
The second and third terms in Eq. \ref{eq26} are equivalent, and the derivation is presented in Section \ref{app1}. The new Hartree energy term can be further divided into two terms:
\begin{equation}\label{eq27}
\begin{gathered}
E_{Hartree}^{new}=\frac{1}{2} \int \rho^{solute}(\boldsymbol{r})\left(\phi^{solute}(\boldsymbol{r})+\phi^{pol}(\boldsymbol{r})\right) d \boldsymbol{r} \\
=\frac{1}{2} \int \rho^{solute}(\boldsymbol{r}) \phi^{solute}(\boldsymbol{r}) d \boldsymbol{r}+\frac{1}{2} \int \rho^{solute}(\boldsymbol{r}) \phi^{pol}(\boldsymbol{r}) d \boldsymbol{r} .
\end{gathered}
\end{equation}
$\phi^{solute}(\boldsymbol{r})$ and $\phi^{p o l}(\boldsymbol{r})$ are the solutions of the vacuum-like Poisson equations for the solute $\rho^{solute}(\boldsymbol{r})$ and polarization $\rho^{pol}(\boldsymbol{r})$ charge densities, respectively\cite{andreussi_revised_2012}.
\subsubsection{Contribution to Kohn-Sham potentials}
\label{subsec2.3.3}
In the original SCCS based on the local solute-solvent interface approach, the potential contribution of $E_{Hartree }^{new,l}$ to the Kohn-Sham potential is as follows\cite{sanchez_first-principles_2009,andreussi_revised_2012}:
\begin{equation}\label{eq28}
V_{Hartree}^{new,l}(\boldsymbol{r})=\phi^{tot,l}(\boldsymbol{r})-\frac{1}{8 \pi} \frac{\partial \epsilon(\boldsymbol{r})}{\partial \rho_{e l}(\boldsymbol{r})}\left|\nabla \phi^{tot,l}(\boldsymbol{r})\right|^2 .
\end{equation}
The symbol '$l$' denotes that the quantity is in the original SCCS approach, which is based on the local solute-solvent interface approach. The derivation of Eq.~\ref{eq28} is provided in Appendix A of Ref.~\citen{sanchez_first-principles_2009}. In contrast, in the solvent-aware algorithm, the dependence of \( \hat{\epsilon}(\boldsymbol{r}) \) on \( \rho_{el}(\boldsymbol{r}) \) is non-local, as outlined in Eqs.~\ref{eqs8}--\ref{eqs12}, and \ref{eqs2}. Consequently, we can rewrite the functional derivative $\frac{\delta \hat{\epsilon}(\boldsymbol{r})_{\left[\rho^{e l}\left(\boldsymbol{r}^{\prime}\right)\right]}}{\delta \rho^{e l}\left(\boldsymbol{r}^{\prime}\right)}$ using the chain rule for functional derivatives as:
\begin{equation}\label{eqs19}
\frac{\delta \hat{\epsilon}(\boldsymbol{r})_{\left[\rho^{e l}\left(\boldsymbol{r}^{\prime}\right)\right]}}{\delta \rho^{e l}\left(\boldsymbol{r}^{\prime}\right)}=\int\left(\int \frac{\delta \hat{\epsilon}(\boldsymbol{r})_{\left[\hat{s}\left(\boldsymbol{r}^{\prime \prime}\right)\right]}}{\delta \hat{s}\left(\boldsymbol{r}^{\prime \prime}\right)} \frac{\delta \hat{s}\left(\boldsymbol{r}^{\prime \prime}\right)_{\left[s\left(\boldsymbol{r}^{\prime \prime \prime}\right)\right]}}{\delta s\left(\boldsymbol{r}^{\prime \prime \prime}\right)} d \boldsymbol{r}^{\prime \prime}\right) \frac{\delta s\left(\boldsymbol{r}^{\prime \prime \prime}\right)}{\delta \rho^{e l}\left(\boldsymbol{r}^{\prime}\right)} d \boldsymbol{r}^{\prime \prime \prime} .
\end{equation}
$\hat{s}\left(\boldsymbol{r}^{\prime \prime}\right)$ non-locally depends on $s\left(\boldsymbol{r}^{\prime \prime\prime}\right)$ according to Eqs. \ref{eqs8}--\ref{eqs10}, and in combination with Eqs. \ref{eqs15} and \ref{eqs13} we can derive $\frac{\delta \hat{s}\left(\boldsymbol{r}^{\prime \prime}\right)_{\left[s\left(\boldsymbol{r}^{\prime \prime \prime}\right)\right]}}{\delta s\left(\boldsymbol{r}^{\prime \prime \prime}\right)}$ as:
\begin{equation}\label{eqs20}
\begin{aligned}
& \frac{\delta \hat{s}\left(\boldsymbol{r}^{\prime \prime}\right)_{\left[s\left(\boldsymbol{r}^{\prime \prime \prime}\right)\right]}}{\delta s\left(\boldsymbol{r}^{\prime \prime \prime}\right)}  \\ & =\frac{\delta\left(s\left(\boldsymbol{r}^{\prime \prime}\right)+\left(1-s\left(\boldsymbol{r}^{\prime \prime}\right)\right) t\left(\boldsymbol{r}^{\prime \prime}\right)_{\left[s\left(\boldsymbol{r}^{\prime \prime \prime}\right)\right]}\right)}{\delta s\left(\boldsymbol{r}^{\prime \prime \prime}\right)} \\ & =\frac{\delta s\left(\boldsymbol{r}^{\prime \prime}\right)}{\delta s\left(\boldsymbol{r}^{\prime \prime \prime}\right)}\left(1-t\left(\boldsymbol{r}^{\prime \prime}\right)_{\left[s\left(\boldsymbol{r}^{\prime \prime \prime}\right)\right]}\right)+\frac{\delta t\left(\boldsymbol{r}^{\prime \prime}\right)_{\left[s\left(\boldsymbol{r}^{\prime \prime \prime}\right)\right]}}{\delta s\left(\boldsymbol{r}^{\prime \prime \prime}\right)}\left(1-s\left(\boldsymbol{r}^{\prime \prime}\right)\right)  \\ & =\left(1-t\left(\boldsymbol{r}^{\prime \prime}\right)\right) \delta\left(\boldsymbol{r}^{\prime \prime \prime}-\boldsymbol{r}^{\prime \prime}\right)+\left(1-s\left(\boldsymbol{r}^{\prime \prime}\right)\right) \frac{\delta t\left(\boldsymbol{r}^{\prime \prime}\right)_{\left[s\left(\boldsymbol{r}^{\prime \prime \prime}\right)\right]}}{\delta s\left(\boldsymbol{r}^{\prime \prime \prime}\right)} \\ & =\left(1-t\left(\boldsymbol{r}^{\prime \prime}\right)\right) \delta\left(\boldsymbol{r}^{\prime \prime \prime}-\boldsymbol{r}^{\prime \prime}\right)+\left(1-s\left(\boldsymbol{r}^{\prime \prime}\right)\right)\left(\int \frac{\delta t\left(\boldsymbol{r}^{\prime \prime}\right)_{[f(\boldsymbol{r})]}}{\delta f(\boldsymbol{r})} \frac{\delta f(\boldsymbol{r})_{\left[s\left(\boldsymbol{r}^{\prime \prime \prime}\right)\right]}}{\delta s\left(\boldsymbol{r}^{\prime \prime \prime}\right)} d \boldsymbol{r}\right) \\ & =\left(1-t\left(\boldsymbol{r}^{\prime \prime}\right)\right) \delta\left(\boldsymbol{r}^{\prime \prime \prime}-\boldsymbol{r}^{\prime \prime}\right) +\left(1-s\left(\boldsymbol{r}^{\prime \prime}\right)\right)\left(\int \frac{d t(f)}{d f}\left(\boldsymbol{r}\right) \delta\left(\boldsymbol{r}^{\prime \prime}-\boldsymbol{r}\right) u\left(\boldsymbol{r}-\boldsymbol{r}^{ \prime \prime \prime}\right) d \boldsymbol{r}\right) \\ & =\left(1-t\left(\boldsymbol{r}^{\prime \prime}\right)\right) \delta\left(\boldsymbol{r}^{\prime \prime \prime}-\boldsymbol{r}^{\prime \prime}\right)+\left(1-s\left(\boldsymbol{r}^{\prime \prime}\right)\right)\left(\frac{d t(f)}{d f}\left(\boldsymbol{r}^{\prime \prime}\right) u\left(\boldsymbol{r}^{\prime \prime}-\boldsymbol{r}^{ \prime \prime \prime}\right)\right). \\ &
\end{aligned}
\end{equation}
From Eq. \ref{eqs12}, one can find:
\begin{equation}\label{eqs21}
\frac{d \hat{\epsilon}(\hat{s})}{d \hat{s}}(\boldsymbol{r})=-\ln \epsilon_0 \hat{\epsilon}(\boldsymbol{r}).
\end{equation}
By applying a similar trick to that used in Eqs.~\ref{eqs5} and \ref{eqs15}, we obtain:
\begin{equation}\label{eqs22}
\frac{\delta \hat{\epsilon}(\boldsymbol{r})_{\left[\hat{s}\left(\boldsymbol{r}^{\prime \prime}\right)\right]}}{\delta \hat{s}\left(\boldsymbol{r}^{\prime \prime}\right)}=\delta\left(\boldsymbol{r}-\boldsymbol{r}^{\prime \prime}\right) \frac{d \hat{\epsilon}(\hat{s})}{d \hat{s}}(\boldsymbol{r}^{\prime \prime}) .
\end{equation}
By combining Eqs.~\ref{eqs22}, \ref{eqs21}, \ref{eqs20}, and \ref{eqs7}, Eq.~\ref{eqs19} can be rewritten as:

\begin{equation}\label{eqs23}
\begin{aligned}
\frac{\delta \hat{\epsilon}(\boldsymbol{r})_{\left[\rho^{e l}\left(\boldsymbol{r}^{\prime}\right)\right]}}{\delta \rho^{e l}\left(\boldsymbol{r}^{\prime}\right)}  &= \int \Bigg[ \int \Big[ -\ln \epsilon_0 \hat{\epsilon}(\boldsymbol{r}) \delta(\boldsymbol{r} - \boldsymbol{r}'') \Big] \\
&\quad \times \Big[ (1 - t(\boldsymbol{r}'')) \delta(\boldsymbol{r}''' - \boldsymbol{r}'') + (1 - s(\boldsymbol{r}'')) \frac{d t(f)}{d f}(\boldsymbol{r}'') u(\boldsymbol{r}''' - \boldsymbol{r}'') \Big] d \boldsymbol{r}'' \Bigg] \\
&\quad \times \delta(\boldsymbol{r}''' - \boldsymbol{r}') \frac{d \epsilon(\rho^{el})}{d \rho^{el}}(\boldsymbol{r}') \frac{1}{\epsilon(\boldsymbol{r}''') \ln \epsilon_0 (-1)} d \boldsymbol{r}''' \\
&= \int \left[ \hat{\epsilon}(\boldsymbol{r}) \left[ (1-t(\boldsymbol{r})) \delta(\boldsymbol{r}'''-\boldsymbol{r}) + (1-s(\boldsymbol{r})) \frac{d t(f)}{d f}(\boldsymbol{r}) u(\boldsymbol{r}'''-\boldsymbol{r}) \right] \right] \\
&\quad \times \delta(\boldsymbol{r}''' - \boldsymbol{r}') \frac{d \epsilon(\rho^{el})}{d \rho^{el}}(\boldsymbol{r}') \frac{1}{\epsilon(\boldsymbol{r}''')} d \boldsymbol{r}''' \\
&= \left[ \hat{\epsilon}(\boldsymbol{r}) \left[ (1-t(\boldsymbol{r})) \delta(\boldsymbol{r}'-\boldsymbol{r}) + (1-s(\boldsymbol{r})) \frac{d t(f)}{d f}(\boldsymbol{r}) u(\boldsymbol{r}'-\boldsymbol{r}) \right] \right] \\
&\quad \times \frac{d \epsilon(\rho^{el})}{d \rho^{el}}(\boldsymbol{r}') \frac{1}{\epsilon(\boldsymbol{r}')}.
\end{aligned}
\end{equation}
The potential contribution of $E_{Hartree}^{new}$ in Eq.~\ref{eq26} to the Kohn-Sham potential is the functional derivative of $E_{Hartree}^{new}$ with respect to the electron density $\rho^{el}(\boldsymbol{r})$:
\begin{equation}\label{eqs24}
V_{Hartree}^{new}(\boldsymbol{r})=\frac{\delta E_{Hartree}^{new}}{\delta \rho^{el}(\boldsymbol{r})}.
\end{equation}
To calculate the functional derivative $\frac{\delta E_{Hartree}^{new}}{\delta \rho^{e l}(\boldsymbol{r})}$, one can utilize the definition of a functional derivative (Eq. A1 in Ref. \citen{sanchez_first-principles_2009}):
\begin{equation}\label{eqs25}
\lim _{\lambda \rightarrow 0} \frac{{{E_{Hartree}^{new}}_{\left[\rho^{el}(\boldsymbol{r})+\lambda f(\boldsymbol{r})\right]}}-{{E_{Hartree}^{new}}_{\left[\rho^{el}(\boldsymbol{r})\right]}}}{\lambda}=\int \frac{\delta E_{Hartree}^{new}}{\delta \rho^{el}(\boldsymbol{r})} f(\boldsymbol{r}) \, d\boldsymbol{r}.
\end{equation}
In the case of the solvent-aware interface, the reformulations of the limit term on the left-hand side of Eq. \ref{eqs25} are the same as those in Eqs. A1-A3 in Ref. \citen{sanchez_first-principles_2009}. Notably, $\epsilon(\boldsymbol{r})$ is now replaced by $\hat{\epsilon}(\boldsymbol{r})$. The reformulation in our case begins to differ starting from the fourth line of Eq. A4 in Ref. \citen{sanchez_first-principles_2009}. For $\epsilon(\boldsymbol{r})$ in Eq. \ref{eqs1}, the following limit can be rewritten using Eq. \ref{eqs6} and the definition of a functional derivative:
\begin{equation}\label{eqs26}
\begin{aligned}
& \lim _{\lambda \rightarrow 0} \frac{\epsilon(\boldsymbol{r})_{\left[\rho^{e l}\left(\boldsymbol{r}^{\prime}\right)+\lambda f\left(\boldsymbol{r}^{\prime}\right)\right]}-\epsilon(\boldsymbol{r})_{\left[\rho^{e l}\left(\boldsymbol{r}^{\prime}\right)\right]}}{\lambda} \\
& =\int \frac{\delta \epsilon(\boldsymbol{r})}{\delta \rho^{e l}\left(\boldsymbol{r}^{\prime}\right)} f\left(\boldsymbol{r}^{\prime}\right) d \boldsymbol{r}^{\prime} \\
& =\int \delta\left(\boldsymbol{r}-\boldsymbol{r}^{\prime}\right) \frac{d \epsilon\left(\rho^{e l}\right)}{d \rho^{e l}}\left(\boldsymbol{r}^{\prime}\right) f\left(\boldsymbol{r}^{\prime}\right) d \boldsymbol{r}^{\prime} \\
& =\frac{d \epsilon\left(\rho^{e l}\right)}{d \rho^{e l}}(\boldsymbol{r}) f(\boldsymbol{r}).
\end{aligned}
\end{equation}
which results in the fourth line of A4 in Ref. \citen{sanchez_first-principles_2009}. However, for $\hat{\epsilon}(\boldsymbol{r})$ in Eq. \ref{eqs12}, according to Eq. \ref{eqs23}, the limit should be rewritten as:
\begin{equation}\label{eqs27}
\begin{gathered}
\lim _{\lambda \rightarrow 0} \frac{\hat{\epsilon}(\boldsymbol{r})_{\left[\rho^{el}\left(\boldsymbol{r}^{\prime}\right)+\lambda f\left(\boldsymbol{r}^{\prime}\right)\right]}-\hat{\epsilon}(\boldsymbol{r})_{\left[\rho^{el}\left(\boldsymbol{r}^{\prime}\right)\right]}}{\lambda} \\
= \int \frac{\delta \hat{\epsilon}(\boldsymbol{r})}{\delta \rho^{el}\left(\boldsymbol{r}^{\prime}\right)} f\left(\boldsymbol{r}^{\prime}\right) \, d\boldsymbol{r}^{\prime} \\
= \int\left(\hat{\epsilon}(\boldsymbol{r}) \left((1-t(\boldsymbol{r})) \delta\left(\boldsymbol{r}^{\prime}-\boldsymbol{r}\right) \right.\right. \\
\left.\left. +(1-s(\boldsymbol{r})) \left(\frac{d t(f)}{d f}(\boldsymbol{r}) u\left(\boldsymbol{r}^{\prime}-\boldsymbol{r}\right)\right)\right)\right) \frac{d \epsilon\left(\rho^{el}\right)}{d \rho^{el}}\left(\boldsymbol{r}^{\prime}\right) \frac{1}{\epsilon\left(\boldsymbol{r}^{\prime}\right)} f\left(\boldsymbol{r}^{\prime}\right) \, d\boldsymbol{r}^{\prime} \\
= \hat{\epsilon}(\boldsymbol{r})(1-t(\boldsymbol{r})) \frac{d \epsilon\left(\rho^{el}\right)}{d \rho^{el}}(\boldsymbol{r}) \frac{1}{\epsilon(\boldsymbol{r})} f(\boldsymbol{r}) \\
+ \int \hat{\epsilon}(\boldsymbol{r})(1-s(\boldsymbol{r})) \frac{d t(f)}{d f}(\boldsymbol{r}) u\left(\boldsymbol{r}^{\prime}-\boldsymbol{r}\right) \frac{d \epsilon\left(\rho^{el}\right)}{d \rho^{el}}\left(\boldsymbol{r}^{\prime}\right) \frac{1}{\epsilon\left(\boldsymbol{r}^{\prime}\right)} f\left(\boldsymbol{r}^{\prime}\right) \, d\boldsymbol{r}^{\prime}
\end{gathered}
\end{equation}
By referring to Eq. A4 in Ref. \citen{sanchez_first-principles_2009} and swapping the order of integration, the limit in Eq. \ref{eqs27} can now be written as:
\begin{equation}\label{eqs28}
\begin{aligned}
& \lim _{\lambda \rightarrow 0} \frac{{{E_{Hartree}^{new}}_{\left[\rho^{el}(\boldsymbol{r})+\lambda f(\boldsymbol{r})\right]}}-{{E_{Hartree}^{new}}_{\left[\rho^{el}(\boldsymbol{r})\right]}}}{\lambda} \\
& = \int \phi^{tot}(\boldsymbol{r}) f(\boldsymbol{r}) \, d\boldsymbol{r} - \frac{1}{8 \pi} \int \left|\nabla \phi^{tot}(\boldsymbol{r})\right|^2 \int \frac{\delta \hat{\epsilon}(\boldsymbol{r})}{\delta \rho^{el}\left(\boldsymbol{r}^{\prime}\right)} f\left(\boldsymbol{r}^{\prime}\right) \, d\boldsymbol{r}^{\prime} \, d\boldsymbol{r} \\
& = \int \phi^{tot}(\boldsymbol{r}) f(\boldsymbol{r}) \, d\boldsymbol{r} \\
& - \frac{1}{8 \pi} \int \left|\nabla \phi^{tot}(\boldsymbol{r})\right|^2 \left[\hat{\epsilon}(\boldsymbol{r})(1-t(\boldsymbol{r})) \frac{d \epsilon\left(\rho^{el}\right)}{d \rho^{el}}(\boldsymbol{r}) \frac{1}{\epsilon(\boldsymbol{r})} f(\boldsymbol{r}) \right. \\
& \left. + \int \hat{\epsilon}(\boldsymbol{r})(1-s(\boldsymbol{r})) \frac{d t(f)}{d f}(\boldsymbol{r}) u\left(\boldsymbol{r}^{\prime}-\boldsymbol{r}\right) \frac{d \epsilon\left(\rho^{el}\right)}{d \rho^{el}}\left(\boldsymbol{r}^{\prime}\right) \frac{1}{\epsilon\left(\boldsymbol{r}^{\prime}\right)} f\left(\boldsymbol{r}^{\prime}\right) \, d\boldsymbol{r}^{\prime}\right] \, d\boldsymbol{r} \\
& = \int \phi^{tot}(\boldsymbol{r}) f(\boldsymbol{r}) \, d\boldsymbol{r} - \frac{1}{8 \pi} \int \left|\nabla \phi^{tot}(\boldsymbol{r})\right|^2 \hat{\epsilon}(\boldsymbol{r})(1-t(\boldsymbol{r})) \frac{d \epsilon\left(\rho^{el}\right)}{d \rho^{el}}(\boldsymbol{r}) \frac{1}{\epsilon(\boldsymbol{r})} f(\boldsymbol{r}) \, d\boldsymbol{r} \\
& - \frac{1}{8 \pi} \int \left[\int \left|\nabla \phi^{tot}\left(\boldsymbol{r}^{\prime}\right)\right|^2 \hat{\epsilon}\left(\boldsymbol{r}^{\prime}\right)\left(1-s\left(\boldsymbol{r}^{\prime}\right)\right) \frac{d t(f)}{d f}\left(\boldsymbol{r}^{\prime}\right) u\left(\boldsymbol{r}-\boldsymbol{r}^{\prime}\right) \, d\boldsymbol{r}^{\prime}\right] \frac{d \epsilon\left(\rho^{el}\right)}{d \rho^{el}}(\boldsymbol{r}) \frac{1}{\epsilon(\boldsymbol{r})} f(\boldsymbol{r}) \, d\boldsymbol{r} \\
& = \int \phi^{tot}(\boldsymbol{r}) f(\boldsymbol{r}) \, d\boldsymbol{r} + \iint -\frac{1}{8 \pi} \left|\nabla \phi^{tot}\left(\boldsymbol{r}^{\prime}\right)\right|^2 \frac{\delta \hat{\epsilon}\left(\boldsymbol{r}^{\prime}\right)}{\delta \rho^{el}(\boldsymbol{r})} \, d\boldsymbol{r}^{\prime} f(\boldsymbol{r}) \, d\boldsymbol{r} .
\end{aligned}
\end{equation}
By the definition of a functional derivative, one can easily see\footnote[3]{To efficiently compute this term within the program, reliance is placed on the fast Fourier transform, as elaborated in Section \ref{subsec3.3}}:
\begin{equation}\label{eqs29}
\begin{aligned}
& V_{Hartree}^{new}(\boldsymbol{r})=\frac{\delta E_{Hartree}^{new}}{\delta \rho^{e l}(\boldsymbol{r})} \\
&=\phi^{tot}(\boldsymbol{r})+\int-\frac{1}{8 \pi}\left|\nabla \phi^{tot}\left(\boldsymbol{r}^{\prime}\right)\right|^2 \frac{\delta \hat{\epsilon}\left(\boldsymbol{r}^{\prime}\right)}{\delta \rho^{e l}(\boldsymbol{r})} d \boldsymbol{r}^{\prime} \\
&=\phi^{tot}(\boldsymbol{r})-\frac{1}{8 \pi}\left|\nabla \phi^{tot}(\boldsymbol{r})\right|^2 \hat{\epsilon}(\boldsymbol{r})(1-t(\boldsymbol{r})) \frac{d \epsilon\left(\rho^{e l}\right)}{d \rho^{e l}}(\boldsymbol{r}) \frac{1}{\epsilon(\boldsymbol{r})} \\
&-\frac{1}{8 \pi}\left[\int\left|\nabla \phi^{t o t}\left(\boldsymbol{r}^{\prime}\right)\right|^2 \hat{\epsilon}\left(\boldsymbol{r}^{\prime}\right)\left(1-s\left(\boldsymbol{r}^{\prime}\right)\right) \frac{d t(f)}{d f}\left(\boldsymbol{r}^{\prime}\right) u\left(\boldsymbol{r}-\boldsymbol{r}^{\prime}\right) d \boldsymbol{r}^{\prime}\right] \frac{d \epsilon\left(\rho^{e l}\right)}{d \rho^{e l}}(\boldsymbol{r}) \frac{1}{\epsilon(\boldsymbol{r})} .
\end{aligned}
\end{equation}
In the solvent-aware case, the potential contribution is given by:
\begin{equation}\label{eq29}
\begin{gathered}
V_{Hartree}^{new}(\boldsymbol{r})=\phi^{tot}(\boldsymbol{r})-\frac{1}{8 \pi}\left|\nabla \phi^{t o t}(\boldsymbol{r})\right|^2\left[-\ln \epsilon_0 \hat{\epsilon}(\boldsymbol{r})\right](1-t(\boldsymbol{r})) \frac{\delta s\left(\rho^{e l}(\boldsymbol{r})\right)}{\delta \rho^{e l}(\boldsymbol{r})} \\
-\frac{1}{8 \pi}\left[\int\left|\nabla \phi^{t o t}\left(\boldsymbol{r}^{\prime}\right)\right|^2\left[-\ln \epsilon_0 \hat{\epsilon}\left(\boldsymbol{r}^{\prime}\right)\right]\left(1-s\left(\boldsymbol{r}^{\prime}\right)\right) \frac{\delta t\left(\boldsymbol{r}^{\prime}\right)}{\delta f\left(\boldsymbol{r}^{\prime}\right)} u\left(\boldsymbol{r}-\boldsymbol{r}^{\prime}\right) d \boldsymbol{r}^{\prime}\right] \frac{\delta s\left(\rho^{e l}(\boldsymbol{r})\right)}{\delta \rho^{e l}(\boldsymbol{r})},
\end{gathered}
\end{equation}
and we found the formula we derived is consistent with the formula provided in Ref. \citen{andreussi_solvent-aware_2019}.\\
\subsubsection{Contribution to analytical forces}
\label{subsec2.3.4}
The contribution of the electrostatic energy $E_{Hartree}^{new}$, as presented in Eq. \ref{eq26}, to the component of the analytical force acting on atom $A$ in direction $\mathscr{i}$, is given by:
\begin{equation}\label{eq30}
f_i^A=-\frac{\partial E_{Hartree}^{new}}{\partial R_i^A}.
\end{equation}
The index $i$ denotes one of the three Cartesian directions of the basis vectors, which represent the positions of the ions. $R_i^A$ represents the projected component of the position vector of nucleus A along the $i$-th direction. By utilizing the product rule, the divergence theorem\footnote[4]{If $\boldsymbol{A}(\boldsymbol{r})$ is a vector field and $a(\boldsymbol{r})$ is a scalar function, then the divergence of the product of $a(\boldsymbol{r})$ and $\boldsymbol{A}(\boldsymbol{r})$ is given by $\nabla \cdot (a(\boldsymbol{r}) \boldsymbol{A}(\boldsymbol{r})) = a(\boldsymbol{r}) \nabla \cdot \boldsymbol{A}(\boldsymbol{r}) + \nabla a(\boldsymbol{r}) \cdot \boldsymbol{A}(\boldsymbol{r})$. Here, $\nabla \cdot$ is the divergence operator, which acts on a vector field such as $\boldsymbol{A}(\boldsymbol{r})$, resulting in $\nabla \cdot \boldsymbol{A}(\boldsymbol{r}) = \frac{\partial A_x}{\partial x} + \frac{\partial A_y}{\partial y} + \frac{\partial A_z}{\partial z}$. $\nabla$ is the gradient operator, which acts on a scalar function like $a(\boldsymbol{r})$, giving $\nabla a(\boldsymbol{r}) = \left(\frac{\partial a}{\partial x} \right) \overrightarrow{\boldsymbol{x}} + \left(\frac{\partial a}{\partial y} \right) \overrightarrow{\boldsymbol{y}} + \left(\frac{\partial a}{\partial z} \right) \overrightarrow{\boldsymbol{z}}$.} and neglecting the surface term, and the generalized Poisson equation in Eq. \ref{eq22}, from the second term in Eq. \ref{eq26} and Eq. \ref{eq30}, the analytical force can be derived as follows:
\begin{equation}\label{eqs30}
\begin{aligned}
f_i^A & =-\frac{\partial E_{Hartree}^{new}}{\partial R_i^A} \\
& =-\frac{\partial \frac{1}{8 \pi} \int \hat{\epsilon}(\boldsymbol{r})\left|\nabla \phi^{t o t}(\boldsymbol{r})\right|^2 d \boldsymbol{r}}{\partial R_i^A} \\
& =-\frac{1}{8 \pi} \int \frac{\partial \hat{\epsilon}(\boldsymbol{r})}{\partial R_i^A}\left|\nabla \phi^{t o t}(\boldsymbol{r})\right|^2 d \boldsymbol{r}-\frac{1}{4 \pi} \int \hat{\epsilon}(\boldsymbol{r}) \nabla \phi^{t o t}(\boldsymbol{r}) \cdot \nabla \frac{\partial \phi^{t o t}(\boldsymbol{r})}{\partial R_i^A} d \boldsymbol{r} \\
& =-\frac{1}{8 \pi} \int \frac{\partial \hat{\epsilon}(\boldsymbol{r})}{\partial R_i^A}\left|\nabla \phi^{t o t}(\boldsymbol{r})\right|^2 d \boldsymbol{r}+\frac{1}{4 \pi} \int \nabla \cdot\left(\hat{\epsilon}(\boldsymbol{r}) \nabla \phi^{t o t}(\boldsymbol{r})\right) \frac{\partial \phi^{t o t}(\boldsymbol{r})}{\partial R_i^A} d \boldsymbol{r} \\
& =-\frac{1}{8 \pi} \int \frac{\partial \hat{\epsilon}(\boldsymbol{r})}{\partial R_i^A}\left|\nabla \phi^{t o t}(\boldsymbol{r})\right|^2 d \boldsymbol{r}-\int \rho^{solute}(\boldsymbol{r}) \frac{\partial \phi^{t o t}(\boldsymbol{r})}{\partial R_i^A} d \boldsymbol{r}
\end{aligned}
\end{equation}
From the third term in Eq. \ref{eq26}, one can also derive the analytical force as follows:
\begin{equation}\label{eqs31}
f_i^A=-\frac{\partial E_{Hartree}^{new}}{\partial R_i^A}=-\frac{1}{2} \int \frac{\partial \rho^{solute}(\boldsymbol{r})}{\partial R_i^A} \phi^{tot}(\boldsymbol{r}) d \boldsymbol{r}-\frac{1}{2} \int \frac{\partial \phi^{tot}(\boldsymbol{r})}{\partial R_i^A} \rho^{solute}(\boldsymbol{r}) d \boldsymbol{r} .
\end{equation}
Since the analytical forces given in Eqs. \ref{eqs30} and \ref{eqs31} are equal, one can establish the following equality:
\begin{equation}\label{eqs32}
\int \rho^{solute}(\boldsymbol{r}) \frac{\partial \phi^{tot}(\boldsymbol{r})}{\partial R_i^A} d \boldsymbol{r}=-\frac{1}{4 \pi} \int \frac{\partial \hat{\epsilon}(\boldsymbol{r})}{\partial R_i^A}\left|\nabla \phi^{tot}(\boldsymbol{r})\right|^2 d \boldsymbol{r}+\int \frac{\partial \rho^{solute}(\boldsymbol{r})}{\partial R_i^A} \phi^{tot}(\boldsymbol{r}) d \boldsymbol{r}.
\end{equation}
By substituting Eq. \ref{eqs32} into Eq. \ref{eqs30}, the form of the analytical force can be rewritten as:
\begin{equation}\label{eqs33}
f_i^A=\frac{1}{8 \pi} \int \frac{\partial \hat{\epsilon}(\boldsymbol{r})}{\partial R_i^A}\left|\nabla \phi^{tot}(\boldsymbol{r})\right|^2 d \boldsymbol{r}-\int \frac{\partial \rho^{solute}(\boldsymbol{r})}{\partial R_i^A} \phi^{tot}(\boldsymbol{r}) d \boldsymbol{r} .
\end{equation}
In the Gaussian and plane wave approach employed in the CP2K software package, the total density ($\rho^{solute}(\boldsymbol{r})$) comprises the total electron density ($\rho^{el}(\boldsymbol{r})$) and the total Gaussian atomic charge density ($\sum_{A} n_c^A(\boldsymbol{r})$)\cite{vandevondele_quickstep_2005}. Here, $n_c^A(\boldsymbol{r})$ represents a Gaussian charge density centered on atom $A$ and corresponds to the charge of the ion described by a norm-conserving Goedecker, Teter, and Hutter (GTH) pseudopotential\cite{goedecker_separable_1996, hartwigsen_relativistic_1998, krack_pseudopotentials_2005}. Since $\hat{\epsilon}(\boldsymbol{r})$ is a functional of $\rho(\boldsymbol{r}^{\prime})$, Eq. \ref{eqs33} can be further derived by swapping the order of integration over $\boldsymbol{r}$ and $\boldsymbol{r}^{\prime}$ and the notations $\boldsymbol{r}$ and $\boldsymbol{r}^{\prime}$:
\begin{equation}\label{eqs34}
\begin{aligned}
f_i^A= & \frac{1}{8 \pi} \int\left(\int \frac{\delta \hat{\epsilon}(\boldsymbol{r})}{\delta \rho^{e l}\left(\boldsymbol{r}^{\prime}\right)} \frac{\partial \rho^{e l}\left(\boldsymbol{r}^{\prime}\right)}{\partial R_i^A} d \boldsymbol{r}^{\prime}\right)\left|\nabla \phi^{tot}(\boldsymbol{r})\right|^2 d \boldsymbol{r}-\int \frac{\partial \rho^{solute}(\boldsymbol{r})}{\partial R_i^A} \phi^{tot}(\boldsymbol{r}) d \boldsymbol{r} \\
= & \frac{1}{8 \pi} \int\left(\int \frac{\delta \hat{\epsilon}\left(\boldsymbol{r}^{\prime}\right)}{\delta \rho^{e l}(\boldsymbol{r})}\left|\nabla \phi^{t o t}\left(\boldsymbol{r}^{\prime}\right)\right|^2 d \boldsymbol{r}^{\prime}\right) \frac{\partial \rho^{e l}(\boldsymbol{r})}{\partial R_i^A} d \boldsymbol{r}-\int \frac{\partial \rho^{e l}(\boldsymbol{r})}{\partial R_i^A} \phi^{t o t}(\boldsymbol{r}) d \boldsymbol{r} \\
& -\int \frac{\partial n_c^A(\boldsymbol{r})}{\partial R_i^A} \phi^{t o t}(\boldsymbol{r}) d \boldsymbol{r} \\
= & -\int\left(-\frac{1}{8 \pi} \int \frac{\delta \hat{\epsilon}\left(\boldsymbol{r}^{\prime}\right)}{\delta \rho^{e l}(\boldsymbol{r})}\left|\nabla \phi^{t o t}\left(\boldsymbol{r}^{\prime}\right)\right|^2 d \boldsymbol{r}^{\prime}+\phi^{t o t}(\boldsymbol{r})\right) \frac{\partial \rho^{e l}(\boldsymbol{r})}{\partial R_i^A} d \boldsymbol{r} \\
& -\int \frac{\partial n_c^A(\boldsymbol{r})}{\partial R_i^A} \phi^{t o t}(\boldsymbol{r}) d \boldsymbol{r} .
\end{aligned}
\end{equation}
According to Eq. \ref{eqs29}, Eq. \ref{eqs34} is as follows:
\begin{equation}\label{eqs35}
f_i^A=-\int V_{Hartree}^{new}(\boldsymbol{r}) \frac{\partial \rho^{el}(\boldsymbol{r})}{\partial R_i^A} d \boldsymbol{r}-\int \frac{\partial n_c^A(\boldsymbol{r})}{\partial R_i^A} \phi^{tot}(\boldsymbol{r}) d \boldsymbol{r}.
\end{equation}
Through our derivation, we have found that the form of Eq. \ref{eqs35} is consistent with the analytical force described in Ref. \citen{andreussi_solvent-aware_2019}. Additionally, the analytical force for the traditional local solute-solvent interface by Andreussi et al.\cite{sanchez_first-principles_2009} is provided in Section \ref{app2} for comparison between the two cases.\\
%%%%%%%%%%%%%%%%%%%%%%%%%%%%%%%%%%%%%%%%%%%%%

\section{Methods}
\label{sec3}
\subsection{Computational setups for DFT}
\label{subsec3.1}
All calculations involved in Section \ref{sec4} were performed using a modified version of the Quickstep module of the CP2K software package\cite{vandevondele_quickstep_2005}. In Quickstep, a mixed Gaussian and plane wave basis is used, for which the Kohn-Sham matrices and the Kohn-Sham orbitals are represented in a Gaussian basis, and the auxiliary plane wave basis is used to solve the Poisson equation in reciprocal space when constructing the Hartree terms in Kohn-Sham matrices to speed up the calculation\cite{vandevondele_quickstep_2005,lippert_hybrid_1997}. The nuclei and some of the non-valence electrons of an atom are effectively treated together as the GTH pseudopotential\cite{goedecker_separable_1996,hartwigsen_relativistic_1998,krack_pseudopotentials_2005}. The basis functions for representing the explicitly treated electrons were of the types of “MOLOPT” (the liquid H\textsubscript{2}O model) which was optimized for molecules, and “MOLOPT-SR” (the rutile TiO\textsubscript{2} and Pt models) which has shorter tails of the radial functions and is suitable for solid phase calculations \cite{vandevondele_gaussian_2007}. The PBE\cite{perdew_generalized_1996} exchange-correlation functional was used. In accordance with the PBE functional, the pseudopotentials used were: GTH-PBE-q1 (hydrogen atoms in the liquid H\textsubscript{2}O models), GTH-PBE-q6 (oxygen atoms in the liquid H\textsubscript{2}O models and the rutile TiO\textsubscript{2} models), GTH-PBE-q12 (Ti atoms), and GTH-PBE-q18 (Pt atoms), and the basis sets used were: DZVP-MOLOPT-GTH-q1 (hydrogen atoms in the liquid H\textsubscript{2}O models), DZVP-MOLOPT-GTH-q6 (oxygen atoms in the liquid H\textsubscript{2}O models), DZVP-MOLOPT-SR-GTH-q6 (oxygen atoms in the rutile TiO\textsubscript{2} models), DZVP-MOLOPT-SR-GTH-q12 (Ti atoms), and DZVP-MOLOPT-SR-GTH-q18 (Pt atoms), respectively. The density cutoff for the auxiliary plane wave basis set, which represents the electron density, was set to 160 Hartree (320 Rydberg) in the calculations with the exception of the calculations of dielectric function isosurfaces, where it was increased to 750 Hartree (1500 Rydberg) to enhance the resolution of the isosurfaces illustrated as figures. The k-point sampling was restricted to the $\Gamma$ point. The convergence threshold of SCF (the EPS\_SCF keyword) was set to 1.0×10\textsuperscript{-10} for all calculations. At the end of each of the converged SCF optimizations, the absolute value of the total energy difference between two adjacent SCF loops was converged to less than 5.0×10\textsuperscript{-7} Hartree.\\
\subsection{Finite difference force in Section \ref{subsec4.1}}
\label{subsec3.2}
To validate the formulas and its implementation of the analytical force given in Eq. \ref{eqs34}, we performed a series of SCCS calculations which were based on the solvent-aware algorithm for the Pt (111) slab. The finite difference atomic forces and the analytical ones acting on a given Pt atom in the Pt (111) slab were calculated. One of the Pt atoms in the outermost layer of the Pt (111) slab was displaced along the $\vec{r}=(1,1,1)$ direction to cause the atom to deviate from its equilibrium position so that it experienced a finite, non-negligible force. The analytical force on this Pt atom after displacement was calculated. Further displacing the atom in the same and opposite directions along the calculated analytical force direction (displacement in the order of $5 \times 10^{-6}$ \r{A}), the numerical force can be calculated from the two displacement sizes with the corresponding two total energy differences. We performed the test on two sets of the $\rho^{min }$ and $\rho^{max }$ parameters (unit of electrons/(Bohr radius)$^3$): {$\rho^{max }=1 \times 10^{-3}$, $\rho^{min }=1 \times 10^{-4}$ } and {$\rho^{max }=1.32 \times 10^{-2}$, $\rho^{min }=5.1 \times 10^{-4}$ } (optimized by fitting the calculated potential of zero charge of the Pt (111) surface to the ab initio result of 4.98 V obtained with explicit water molecules as the solvent\cite{sakong_structure_2016}, and the experimental capacitances of a series of metal/water interfaces\cite{hormann_grand_2019}). We also calculated the analytical and numerical forces by using the SCCS model which was already implemented and tested in CP2K\cite{yin_periodic_2017} based on the normal local solvent-solute interface proposed by Andreussi et al.\cite{andreussi_revised_2012} as a comparison.\\
\subsection{Efficient convolution calculations in the implementation}
\label{subsec3.3}
The direct computation of convolution in real space is usually very expensive, as it involves performing a full-space integral calculation at each point in space. In Eqs. \ref{eqs10} and \ref{eqs29}, calculations involving real-space convolution are needed. In the implementation of the solvent-aware algorithm in CP2K, we transform the aforementioned convolution calculations into the product of two integrand functions in reciprocal space. The functions to be integrated are represented as periodic real-space grids in our implementation. These real-space grids are transformed into reciprocal space using fast Fourier transforms and the function multiplication is performed. The product of the two functions is then transformed back into real space using fast Fourier transforms\footnote[5]{The convolution theorem states that the Fourier transform of the convolution of functions is the product of their Fourier transforms.}.
\section{Results and discussion}
\label{sec4}
\subsection{Analytical force validation}
\label{subsec4.1}
\begin{figure}[H]
\centering
\includegraphics[width=10cm]{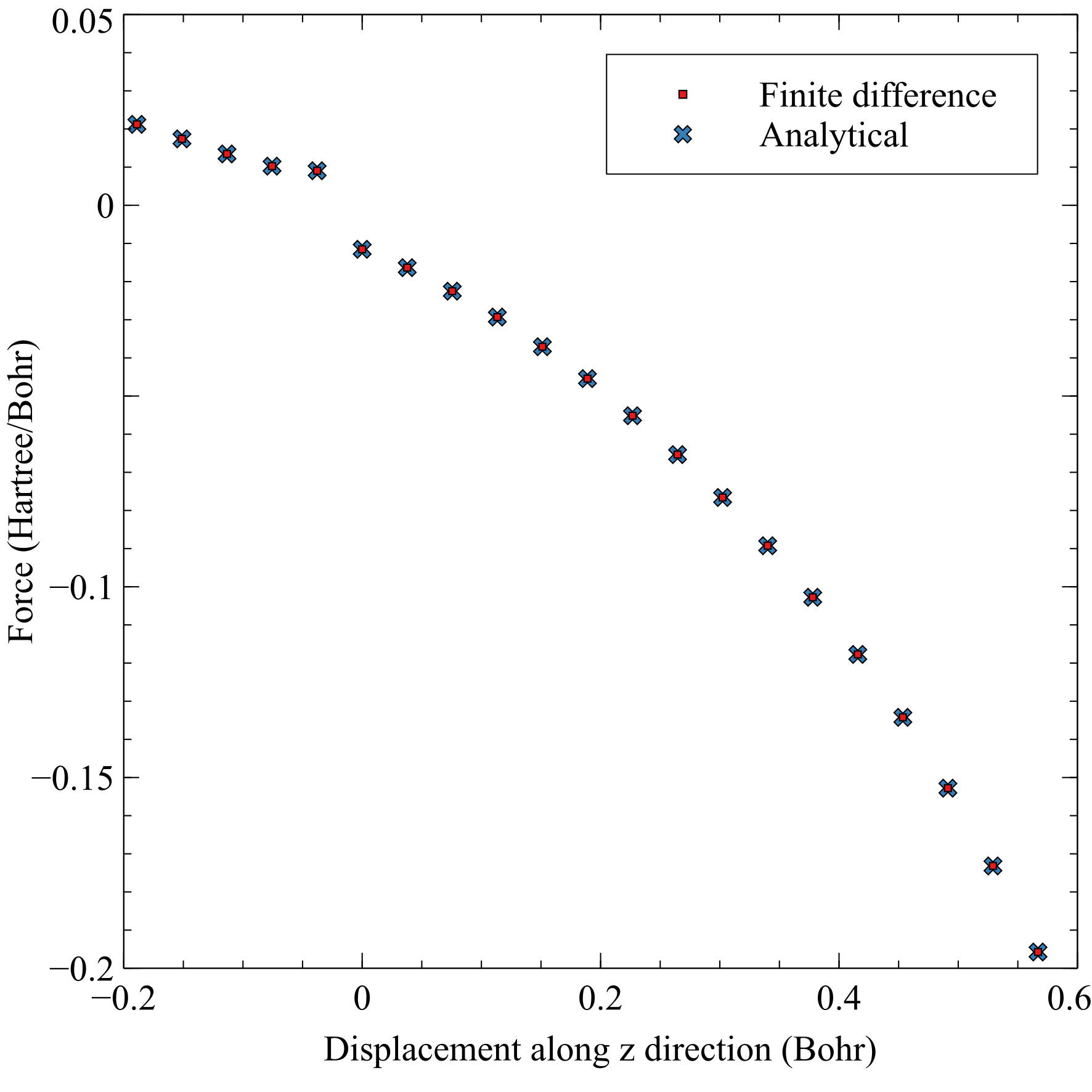}
\caption{Total atomic forces act on the Pt atom in the outermost layer of the Pt (111) surface when displaced along the \(\vec{r}=(1,1,1)\) direction. Small square signs and cross signs represent finite difference forces and analytical forces, respectively. These forces were calculated using DFT in combination with our new SCCS implementation based on the solvent-aware interface. The solute-solvent boundary parameters used are: \(\{\rho^{max}=1 \times 10^{-3}, \rho^{min}=1 \times 10^{-4}\}\).}\label{fig2}
\end{figure}

In this section, we present the analytical and numerical forces calculated for the Pt(111) surface model using DFT+SCCS. The analytical forces were calculated directly inside the program, while the numerical forces were calculated using the finite difference method, which is the response of the total system energy to perturbations in the position of one Pt atom. Fig. \ref{fig2} presents the analytical and numerical forces acting on the Pt atom in the outermost layer of the Pt slab at various position displacements, ranging from -0.19 Bohr to 0.57 Bohr. These forces were calculated using DFT in combination with our new SCCS implementation based on the solvent-aware interface. The solute-solvent boundary parameters used were the commonly applied ones: $\rho^{max}=1 \times 10^{-3}$, $\rho^{min}=1 \times 10^{-4}$ (units: electrons/(Bohr radius)$^3$). As shown in Fig. \ref{fig2}, the analytical forces and the corresponding finite difference forces are closely matched across the range from -0.19 to 0.57 Bohr for displacements. The differences between the analytical and numerical forces are summarized in Table~\ref{tabs1a}. It is evident from Table~\ref{tabs1a} that the maximum deviation between the analytical and numerical forces is on the order of \(1.0 \times 10^{-7}\) Hartree/Bohr and there is a high degree of agreement between the analytical and finite difference forces.\\
The solute-solvent boundary parameters \(\{\rho^{max}=1.32 \times 10^{-2}, \rho^{min}=5.1 \times 10^{-4}\}\) were found to be optimal for reproducing both the calculated \textit{ab initio} potential of zero charge (PZC) for the Pt (111) surface using explicit H\(_2\)O molecules as the solvent molecules\cite{sakong_structure_2016} and the experimental capacitances of a series of metal/water interfaces\cite{hormann_grand_2019}. The optimization process will be discussed in detail in another separate paper in the future. It is therefore necessary to evaluate the deviation between the analytical and finite difference forces under these optimal parameters. Fig.~\ref{fig2} presents the analytical and numerical forces calculated using DFT combined with our new SCCS implementation based on the solvent-aware interface, and utilizing the parameters \(\{\rho^{max}=1.32 \times 10^{-2}, \rho^{min}=5.1 \times 10^{-4}\}\). As shown in Fig. \ref{figs3}\subref{figs3a}, the analytical forces and the corresponding finite difference forces match up nicely across the range from -0.19 to 0.57 Bohr displacements. The differences between the analytical and numerical forces are summarized in Table \ref{tabs1b}. One can see from Table~\ref{tabs1b} that the maximum deviation between the analytical and finite difference forces is on the order of \(1.0 \times 10^{-4}\) Hartree/Bohr, which is three orders of magnitude higher than that calculated using the previous parameter set. Due to this significant increase in maximum deviation when transitioning from lower- to higher-value solute-solvent boundary parameters, we conducted tests using DFT+SCCS based on the local solute-solvent boundary defined by Andreussi et al. (Eqs.~\ref{eqs1}--\ref{eqs3} in this paper)\cite{yin_periodic_2017}. The analytical and finite difference forces calculated using the parameters \(\{\rho^{max}=1 \times 10^{-3}, \rho^{min}=1 \times 10^{-4}\}\) and \(\{\rho^{max}=1.32 \times 10^{-2}, \rho^{min}=5.1 \times 10^{-4}\}\) are shown in Figs. \ref{figs3}\subref{figs3b} and \ref{figs3}\subref{figs3c}, respectively. The differences between the analytical forces and the corresponding finite difference forces are summarized in Table~\ref{tabs1c} and Table~\ref{tabs1d}, respectively. As shown in Figs.~\ref{figs3b} and~\ref{figs3c}, the analytical and finite difference forces closely match. The maximum deviation between these forces is on the order of \(1.0 \times 10^{-7}\) Hartree/Bohr for the parameters \(\{\rho^{max} = 1 \times 10^{-3}, \rho^{min} = 1 \times 10^{-4}\}\), as detailed in Table~\ref{tabs1c}. For the parameters \(\{\rho^{max} = 1.32 \times 10^{-2}, \rho^{min} = 5.1 \times 10^{-4}\}\), the maximum deviation increases significantly to \(1.0 \times 10^{-3}\) Hartree/Bohr, which is four orders of magnitude higher, as shown in Table~\ref{tabs1d}. These results suggest that for the standard implementation of SCCS based on the Andreussi et al.'s local solute-solvent boundary within CP2K, larger boundary parameters lead to a rapid increase in deviation. It is also noteworthy that the signs of the calculated force differences, as reported in Table~\ref{tabs1b}, remain consistent when using the SCCS model based on the solvent-aware algorithm. Further studies are necessary to explore these observations.\\
\subsection{Tests for solute-solvent boundary and SCF convergence}
\label{subsec4.2}
\subsubsection{The solute-solvent boundaries in local and non-local interface algorithms}
\label{subsec4.2.1}
\begin{figure}[H]
\centering
\subfigure[]{
\includegraphics[height=0.153666\textwidth]{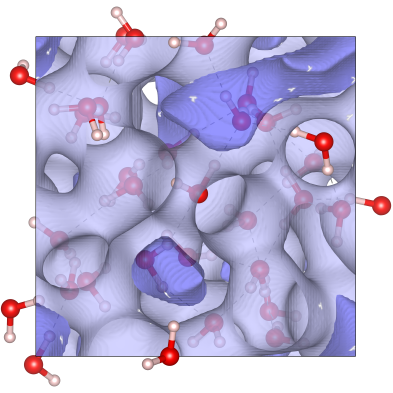}
\label{fig1a}
}
\hspace{0pt}
\subfigure[]{
\includegraphics[width=0.153666\textwidth, angle=90]{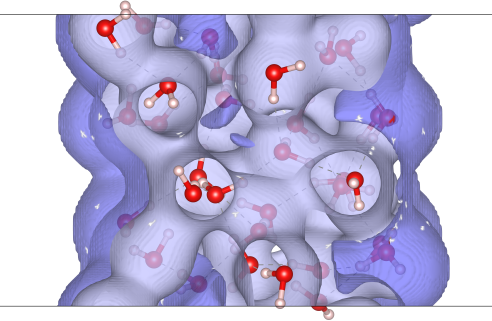}
\label{fig1c}
}
\hspace{0pt}
\subfigure[]{
\includegraphics[height=0.153666\textwidth]{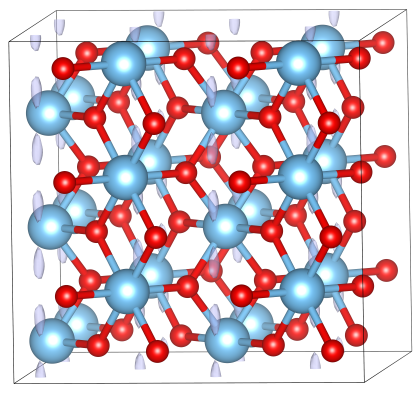}
\label{fig1e}
}
\hspace{0pt}
\subfigure[]{
\includegraphics[width=0.153666\textwidth, angle=90]{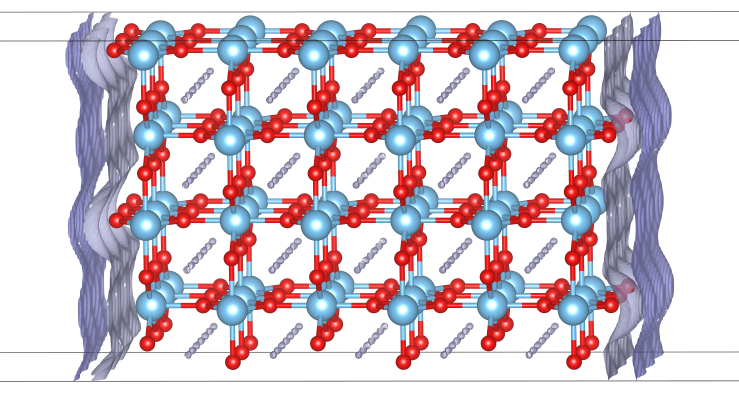}
\label{fig1g}
}
\hspace{0pt}
\subfigure[]{
\includegraphics[width=0.153666\textwidth, angle=90]{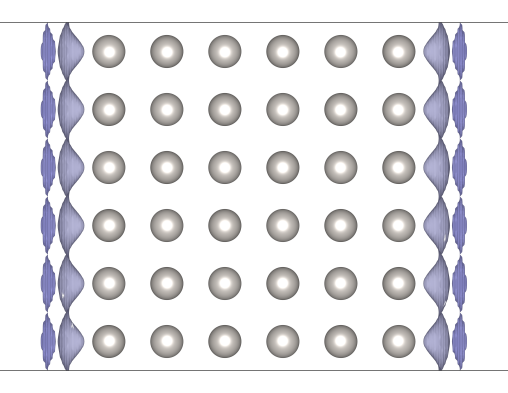}
\label{fig1i}
}
\hspace{0pt}
\subfigure[]{
\includegraphics[width=0.153666\textwidth, angle=90]{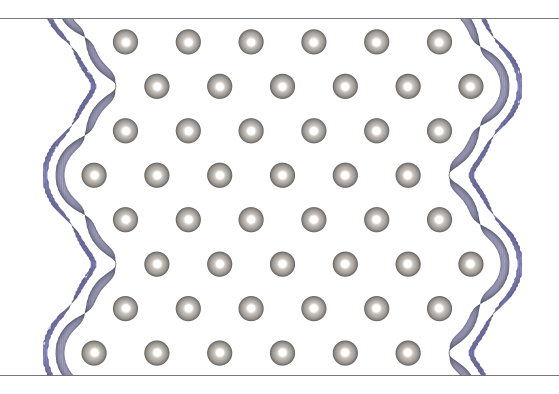}
\label{fig1k}
}
\hspace{0pt}
\subfigure[]{
\includegraphics[height=0.153666\textwidth]{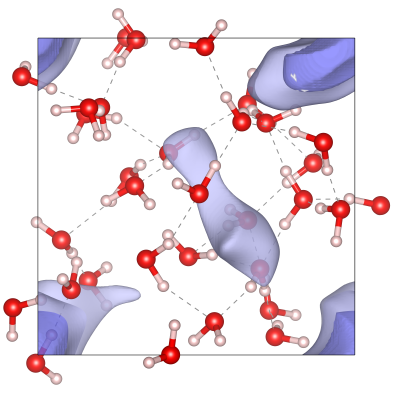}
\label{fig1b}
}
\hspace{0pt}
\subfigure[]{
\includegraphics[width=0.153666\textwidth, angle=90]{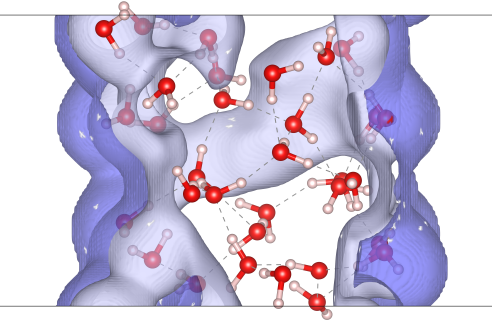}
\label{fig1d}
}
\hspace{0pt}
\subfigure[]{
\includegraphics[height=0.153666\textwidth]{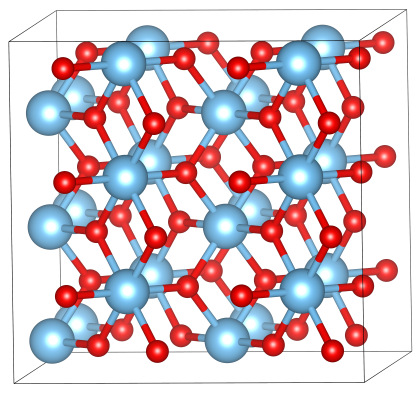}
\label{fig1f}
}
\hspace{0pt}
\subfigure[]{
\includegraphics[width=0.153666\textwidth, angle=90]{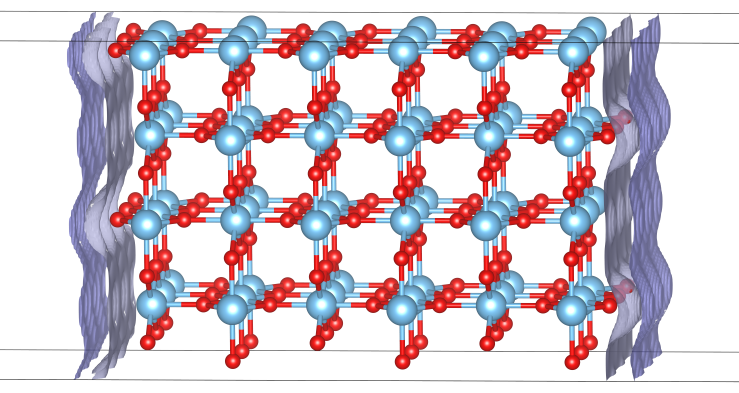}
\label{fig1h}
}
\hspace{0pt}
\subfigure[]{
\includegraphics[width=0.153666\textwidth, angle=90]{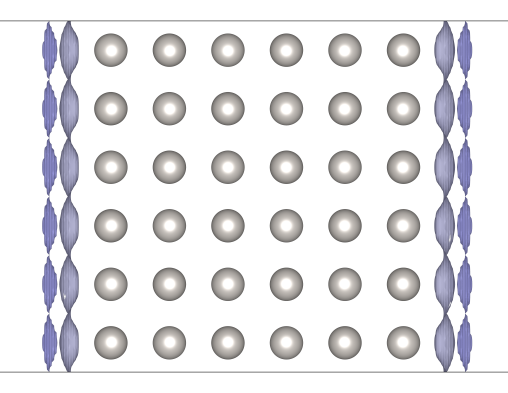}
\label{fig1j}
}
\hspace{0pt}
\subfigure[]{
\includegraphics[width=0.153666\textwidth, angle=90]{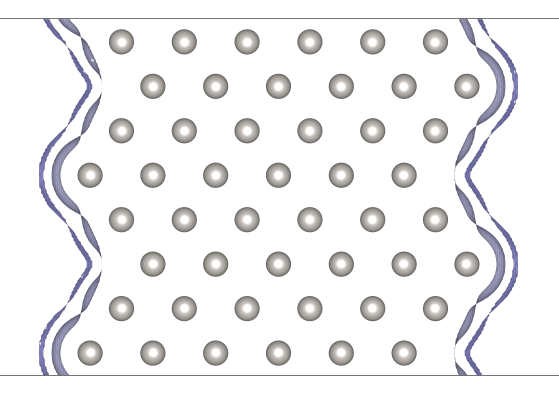}
\label{fig1l}
}
\quad
\caption{
Isosurfaces of the dielectric functions calculated using the traditional Andreussi et al.'s local solute-solvent interface algorithm or the newly implemented solvent-aware algorithm. In the figures, deep blue and light blue represent isosurface levels of 78.2 and 1.1, respectively, except for sub-figures (c), (d), and (j) which use 1.01 to make the contours more easily discernible. The upper half of the figure presents results from the old local interface implementation (sub-figures (a)-(f)), while the lower half shows results from the solvent-aware algorithm implementation (sub-figures (g)-(l)). From top to bottom, each pair of sub-figures corresponds to the same test system: liquid H\(_2\)O bulk ((a) and (g)), liquid H\(_2\)O surface ((b) and (h)), rutile TiO\textsubscript{2} bulk ((c) and (i)), (110) rutile TiO\textsubscript{2} surface ((d) and (j)), (100) surface of Pt ((e) and (k)), and the ``missing row'' reconstructed (110) surface of Pt ((f) and (l)). All systems were calculated using $\rho^{max} = 1.32 \times 10^{-2}$ and $\rho^{min} = 5.1 \times 10^{-4}$ as density boundary values. In the case of the Pt bulk system, the dielectric functions are always 1 throughout the simulation cell, and their isosurfaces are not shown in this figure.
}\label{fig1}
\end{figure}

After demonstrating the consistency of the newly implemented SCCS energy and analytical forces, we now focus on the computational results of the solute-solvent interface provided by this implementation. Initially, we compare the solute-solvent interfaces from the new solvent-aware algorithm implementation and the older local interface implementation under identical electronic densities. Our tests cover various system types, including bulk and surface models, namely liquid H\(_2\)O, and bulk rutile TiO\(_2\) and Pt, all commonly used in DFT calculations. The bulk model of liquid H\(_2\)O was selected randomly from a pre-equilibrated trajectory of a 300K NVT DFT-MD simulation, matching the density of room temperature liquid H\(_2\)O. The bulk models of rutile TiO\(_2\) and Pt were constructed using optimized cell parameters and atomic coordinates by using PBE. We cleaved and built the 6-layer (2 \(\times\) 3) (110) surface for rutile TiO\(_2\), the 6-layer (3 \(\times\) 3) (100) surface for Pt, and the 6-layer (2 \(\times\) 6) “missing row” reconstructed (110) surface for Pt from these bulk structures. The computational settings and parameters remained consistent across tests unless modifications were explicitly mentioned. We maintained the density thresholds \(\rho^{max} = 1 \times 10^{-2}\), \(\rho^{min} = 1 \times 10^{-3}\), which are of the same order of magnitude as those used in previous studies of metal slab systems~\cite{hormann_grand_2019}. The parameters for the solvent-aware algorithm were set to recommended values from the original literature \(\left(R_{solv}=2.6, \alpha_\zeta=2, \Delta_\zeta=0.5, f_0=0.65, \Delta_\eta=0.02\right)\), which are suitable for using implicit solvent of liquid H\(_2\)O\cite{andreussi_solvent-aware_2019}.\\
The interface between solute and solvent can be characterized by the dielectric function ($\hat{\epsilon}(\boldsymbol{r})$ or $\epsilon(\boldsymbol{r})$), which is determined from the interface function ($\hat{s}(\boldsymbol{r})$ and $s(\boldsymbol{r})$) as given in Eqs.\ref{eqs12} and \ref{eqs1}, respectively. The isosurfaces of the calculated dielectric functions are displayed in Figs.~\ref{fig1a} and~\ref{fig1b} for liquid H\(_2\)O bulk, Figs.~\ref{fig1c} and~\ref{fig1d} for the liquid H\(_2\)O surface, Figs.~\ref{fig1e} and~\ref{fig1f} for rutile TiO\(_2\) bulk, Figs.~\ref{fig1g} and~\ref{fig1h} for the (110) rutile TiO\(_2\) surface, Figs.~\ref{fig1i} and~\ref{fig1j} for the Pt (100) surface, and Figs.~\ref{fig1k} and~\ref{fig1l} for the Pt “missing row” reconstructed (110) surface. Figs.~\ref{fig1}\subref{fig1a}--\subref{fig1k} and \ref{fig1}\subref{fig1b}--\subref{fig1l} show the results calculated using the traditional Andreussi et al.'s local interface algorithm and the non-local solvent-aware interface algorithm, respectively. Light blue and dark blue represent isosurface levels of 1.1 and 78.2, respectively, denoting positions adjacent to pure solute (with a dielectric function value of 1) and adjacent to pure implicit water (with a dielectric function value of 78.3) in the solute-solvent transition region.\\
Fig.~\ref{fig1a} shows the computational results of the traditional Andreussi et al.'s local solute-solvent interface algorithm for the liquid H\(_2\)O bulk system\cite{andreussi_revised_2012}. In spatial regions closest to H\(_2\)O molecules and their hydrogen bonds, the high electron density values prevent any implicit solvent from completely or partially occupying these regions under the current computational setup. In the other cavity regions, the electronic densities are relatively low, so these areas are filled with implicit solvent to varying degrees. Fig.~\ref{fig1b} displays the distribution of the dielectric function obtained using a non-local solvent-aware interface calculation. It can be seen that most of the regions previously mentioned, which are occupied by implicit water, do not contain implicit solvent and are in a state without implicit solvent. This result aligns with our expectations for the solvent-aware algorithm. Figs.~\ref{fig1c} and \ref{fig1d} show the isosurfaces of the dielectric functions predicted by the two interface algorithms in the case of the liquid H\(_2\)O surface model. The comparison between the two figures reveals that within the interior of the explicit water model, the solvent-aware algorithm significantly reduces the presence of implicit water, while the interface between the implicit water and the transition region remains almost unchanged. This occurs because, for positions primarily surrounded by pure implicit solvent within the detection sphere (with a radius of $\alpha_\zeta R_{solv} = 5.2$ a.u.)\cite{andreussi_solvent-aware_2019}, the solvent-aware algorithm strives to preserve the dielectric function provided by the original local algorithm. Conversely, if a position is less occupied by implicit solvent, the algorithm tends to set the dielectric function to 1.\\
Figs.~\ref{fig1e}, \ref{fig1f} and Figs.~\ref{fig1g}, \ref{fig1h} display the test results for rutile TiO\textsubscript{2} bulk and its (110) surface, respectively. One can observe from Fig. \ref{fig1e} that many small spindle-shaped regions with dielectric functions greater than 1 appear within the TiO\textsubscript{2} bulk. These regions are physically unreasonable because, in principle, the interior of a periodic titanium dioxide bulk should not be occupied by any implicit solvent. Tests on the convergence of Kohn-Sham SCF also demonstrate that these non-physical dielectric function regions lead to SCF convergence disasters (as shown in Table \ref{fig4}). In the scenario of the solvent-aware algorithm, these incorrect implicit solvent regions can be entirely eliminated, resulting in a calculated dielectric function of 1 throughout the entire simulation cell. In calculations of the TiO\textsubscript{2} (110) surface using traditional Andreussi et al.'s local interface algorithms, many spindle-shaped implicit solvent regions again emerge within the TiO\textsubscript{2} slab (Fig.~\ref{fig1g}). Consistent with the previous tests, the solvent-aware algorithm continues to eliminate the non-physical, incorrect implicit solvent regions within the solid and does not obviously influence the boundaries of the pure solvent region (dark blue isosurfaces). In addition, as expected, the small bump-like portions of the light blue isosurfaces, which are adjacent to the TiO\textsubscript{2} slab and extend towards it, have been smoothed out. The same phenomenon is also observed in the calculation results for the (100) surface and the “missing row” reconstructed (110) surface of Pt, as shown in Figs. \ref{fig1i} and \ref{fig1j}, and Figs. \ref{fig1k} and \ref{fig1l}, respectively. It should be noted that with the current computational settings, for both types of solute-solvent interface algorithms, the calculated dielectric function values within the Pt bulk and the deeper regions of the Pt slabs are consistently equal to 1.\\

%\afterpage{
%\clearpage
%}
\subsubsection{Influences of the \(f_0\) and \(R_{solv}\) parameters on solute-solvent boundaries}
\label{subsec4.2.2}
\begin{figure}[H]
\centering
\subfigure[]{
\includegraphics[width=\textwidth]{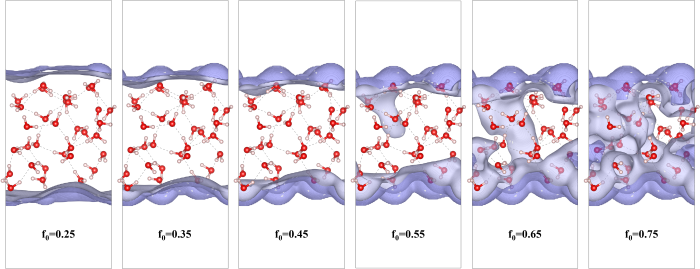}
\label{fig3a}
}
\subfigure[]{
\includegraphics[width=\textwidth]{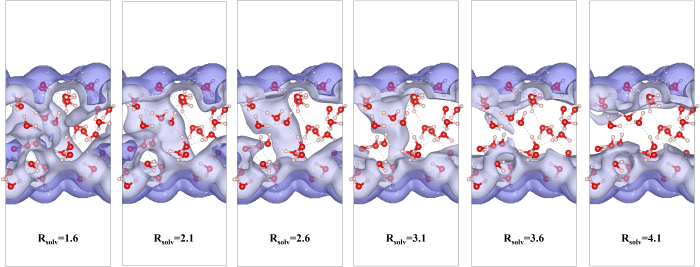}
\label{fig3b}
}
\quad
\caption{
Isosurfaces of the dielectric functions for the H\(_2\)O surface system calculated using the newly implemented solvent-aware algorithm: (a) at the fixed R\textsubscript{solv}=2.6 a.u. and various f\textsubscript{0} values (0.25-0.75), (b) at the fixed f\textsubscript{0}=0.65 and various R\textsubscript{solv} values (1.6-4.1 a.u.), where deep blue and light blue represent isosurface levels of 78.2 and 1.1, respectively. In all cases, $\rho^{max }=1.32 \times 10^{-2}$, and $\rho^{min }=5.1 \times 10^{-4}$ density boundary values were consistently used.}\label{fig3}
\end{figure}

Next, we test the impact of the \(f_0\) and \(R_{solv}\) parameters (see Eqs. \ref{eqs9} and \ref{eqs11}, respectively) in the solvent-aware algorithm on the calculation results of the solute-solvent interface (dielectric function). The H\(_2\)O surface model, with its complex interface shape and transition region, was chosen as the test system. The computational settings and parameters remained unchanged from previous tests, except for adjustments to \(f_0\) or \(R_{solv}\). The \(f_0\) parameter can be qualitatively understood as the critical value for the proportion of the pure solute region within the detection sphere. When the proportion of pure solute inside the detection sphere exceeds \(f_0\), the dielectric function at the sphere's center rapidly approaches 1. Conversely, if this proportion falls below \(f_0\), the dielectric function quickly reverts to that predicted by the local interface approach. Thus, an excessively high \(f_0\) can hinder the elimination of implicit solvent regions encroaching into the solute area. When \(f_0\) exceeds 1 (the value when the detection sphere is entirely occupied by pure solute), the solvent-aware interface gradually reverts to the conventional local interface by Andreussi et al., denoted as \(\epsilon(\boldsymbol{r})\). According to Ref.~\citen{andreussi_solvent-aware_2019}, all points within a spherical region of radius \(R_{solv}\) surrounded by solute, and those within a cylindrical region of pure solvent with a cross-sectional radius of \(R_{solv}\) (where the dielectric function equals 78.3 for liquid H\(_2\)O), can be identified by the solvent-aware algorithm as being within the solute region, setting their dielectric functions to 1. This is achievable only if \(f_0 \leq 0.65\) (when \(\alpha_\xi = 2\)).\\
Based on the same geometric structure and electronic density of the H\(_2\)O surface system, Fig.~\ref{fig3a} illustrates the isosurfaces of the dielectric functions calculated using our CP2K implementation of the solvent-aware algorithm at various \(f_0\) parameters. The \(R_{solv}\) parameter was fixed at 2.6 a.u., while the \(f_0\) parameter gradually increased from 0.25 to 0.75. As demonstrated in Fig.~\ref{fig3a}, consistent with the analysis above, an increase in the \(f_0\) parameter results in greater retention of implicit solvent within the H\(_2\)O slab. Furthermore, it is known that an \(f_0\) parameter less than 0.5 shifts the perfectly planar interface between pure implicit solvent and pure solute towards the pure solvent region, as analyzed in Ref.~\citen{andreussi_solvent-aware_2019}. This shift occurs because positions near the interface within the pure implicit solvent region may have \(f(x)\) values greater than \(f_0\) and less than 0.5. As the positive difference \(f(x) - f_0\) increases, the dielectric function value at these positions predicted by the solvent-aware algorithm rapidly approaches 1. This indicates that as the \(f_0\) value decreases, the boundary gradually moves towards the solvent region. If \(f_0\) is greater than 0.5, then the \(f(x)\) values across the entire pure solvent region (all less than 0.5) are smaller than \(f_0\). This means that the dielectric functions predicted by the traditional Andreussi et al.'s local solute-solvent interface algorithm within the pure solvent region do not change obviously by the solvent-aware algorithm. Consistently, Fig.~\ref{fig3a} shows that when \(f_0\) exceeds 0.5, the boundary of the pure solvent region exhibits no noticeable changes as \(f_0\) increases from 0.55 to 0.75. However, when \(f_0\) is less than 0.5, the boundary of the pure solvent region continuously moves inward as \(f_0\) decreases. Furthermore, as \(f_0\) decreases from 0.45 to 0.25, the boundary of the pure solvent region becomes progressively smoother due to the planar boundary of the pure solvent gradually expanding inward in a translational manner, causing the originally spherical concave areas of the pure solvent boundary to become part of the pure solute region.\\
%\afterpage{
%\clearpage
%}
In the solvent-aware algorithm, \(R_{solv}\) possesses the physical meaning of solvent size in a statistical sense which is the basis on which the parameter is chosen, and along with \(\alpha_{\xi}\), determines the effective radius (\(R_{solv}\alpha_{\xi}\)) of the detection sphere. In practical calculations targeting the microscopic configurations of the liquid H\(_2\)O surface (such as the geometry structure in a specific microscopic state of the system), \(R_{solv}\) representing the size of H\(_2\)O molecules (e.g., \(R_{solv}=2.6\) a.u., consistent with the experimental O-O pair correlation function) does not necessarily have to perfectly eliminate the implicit solvent regions within the liquid H\(_2\)O slab. From the perspective of algorithm design, a too small radius of the detection sphere will render it incapable of effectively accommodating larger-sized implicit solvent cavities or pockets, leading to a too low value of \(f(x)\) at the sphere's center, thereby preventing the dielectric function at that position from being modified by the solvent-aware algorithm. It is foreseeable that eliminating larger-sized implicit solvent regions requires a larger \(R_{solv}\) value. As illustrated in Fig.~\ref{fig3b}, as \(R_{solv}\) increases from 1.6 a.u.to 4.6 a.u., the invasion of the implicit solvent into the H\(_2\)O slab is greatly alleviated. When \(R_{solv}\) is less than or equal to 3.1 a.u., the implicit solvent regions penetrating the H\(_2\)O slab cannot be completely eliminated. However, when \(R_{solv}\) increases to 4.1 a.u., such regions are effectively eliminated. At the same time, one can observe that the boundary of the pure solvent is almost unaffected by changes in the \(R_{solv}\) parameter. The reason is that, when \(f_0\) is greater than 0.5, the \(f(x)\) values in the pure solvent region are generally smaller than \(f_0\), and the dielectric function values will not be modified by the solvent-aware algorithm.\\
\subsubsection{Improvement of the SCF convergence for DFT+SCCS calculations}
\label{subsec4.2.3}
\begin{figure}[H]
\centering

\includegraphics[width=\textwidth]{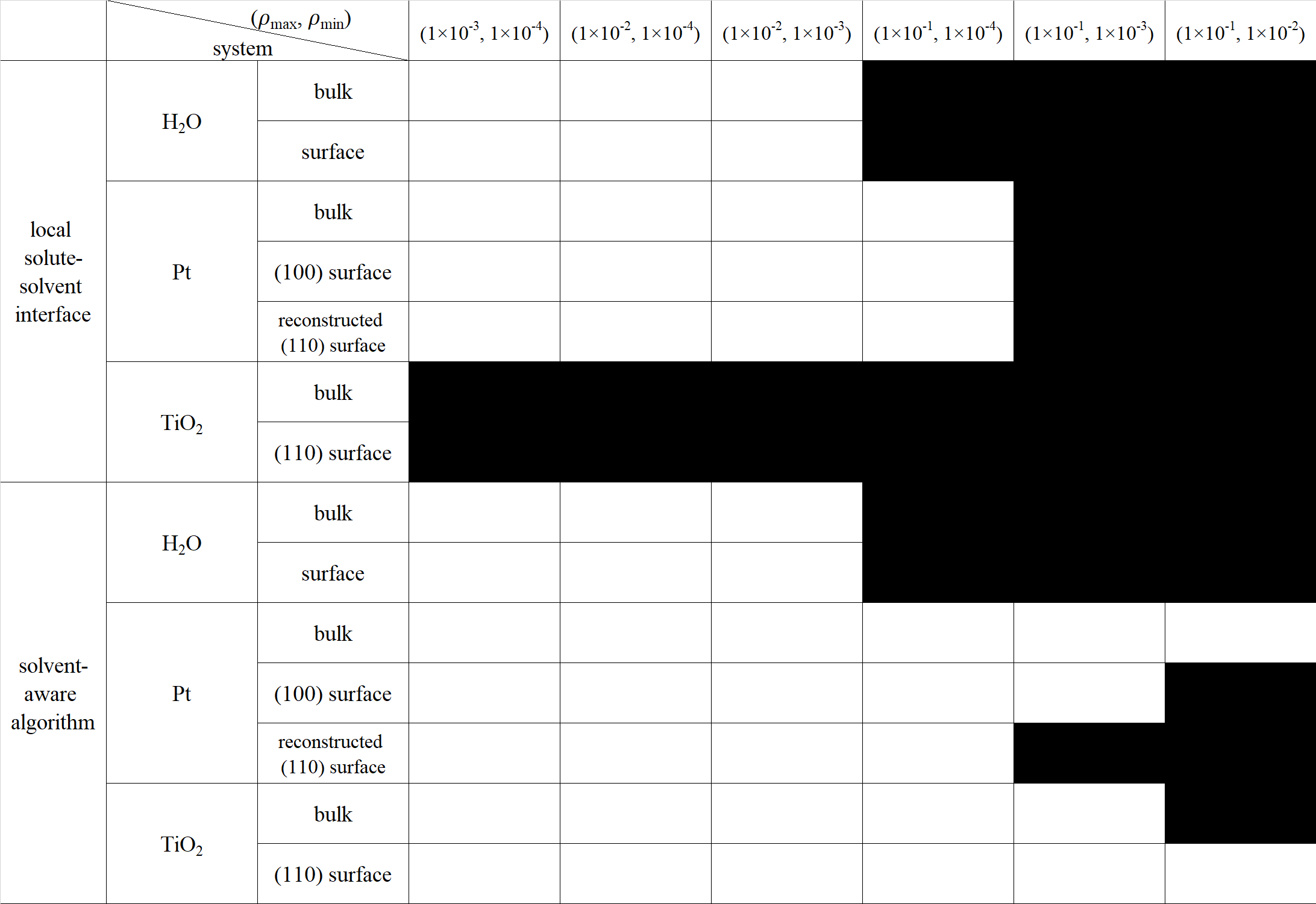}
\captionsetup{name=Table}
\renewcommand{\thefigure}{1}
\caption{The SCF convergence of DFT+SCCS calculations within 500 iterations under various $\rho^{max}$ and $\rho^{min}$ parameters, using both the old CP2K SCCS implementation based on the Andreussi et al.'s local solute-solvent interface approach and our CP2K implementation of the modified SCCS based on the non-local solvent-aware solute-solvent interfaces. The diagonalization approach and the modified Pulay mixing approach were used in all calculations.  In the table, white and black indicate the convergence and divergence of SCF iterations within 500 steps, respectively.}
\renewcommand{\thefigure}{\arabic{figure}}
\label{fig4}
\end{figure}

Excessively high density threshold parameters, \(\{\rho^{max}, \rho^{min}\}\), may lead to unreasonable solute-solvent interfaces and thus cause disasters in SCCS DFT's SCF convergence. The severity of convergence issues varies across different types of systems. Table~\ref{fig4} presents the success (white blocks) and failure (black blocks) of SCF convergence within 500 iterations of DFT+SCCS calculations for the systems shown in Fig.~\ref{fig1}, using a range of \(\rho^{max}\) and \(\rho^{min}\) parameters. These calculations employ the old CP2K implementation based on the traditional Andreussi et al.'s local solute-solvent interfaces and our CP2K implementation of the modified SCCS approach based on the non-local solvent-aware solute-solvent interfaces, respectively. Each of the chosen \(\rho^{max}\) and \(\rho^{min}\) parameters spans three orders of magnitude, with \(\rho^{max}\) always being greater than \(\rho^{min}\). All SCF optimizations in Table~\ref{fig4} are based on an improved Pulay density mixing method\cite{sundararaman_grand_2017} and employ the same density mixing parameters. SCF optimization methods also impact SCF convergence, with test results based on the Broyden density mixing method and the orbital transformation electron density optimization method\cite{VandeVondele_ot_2003,weber_ot_2008} presented separately in Tables~\ref{figs3} and \ref{figs4}.\\
From Table~\ref{fig4}, it is evident that with the gradual increase in the $\rho^{max}$ and $\rho^{min}$ parameters, the convergence of SCF generally tends to deteriorate. For systems such as the rutile TiO\textsubscript{2} bulk, the (110) rutile TiO\textsubscript{2} surface, the Pt bulk, and the Pt (100) surface, systematically improved SCF convergence is observed when using our CP2K implementation of the solvent-aware algorithm. Notably, for the bulk and surface of TiO\textsubscript{2} when the calculations were performed using the old SCCS implementation\cite{yin_periodic_2017} based on the Andreussi et al.'s local solute-solvent interface approach, all SCFs fail to converge within 500 steps. However, when employing the solvent-aware algorithm, except for calculations on bulk TiO\textsubscript{2} with significantly high $\rho^{max} = 1 \times 10^{-1}$ and $\rho^{min} = 1 \times 10^{-2}$, the remaining SCFs can successfully converge. Based on the tested results reported in Ref.~\citen{andreussi_solvent-aware_2019} and our calculations for other semiconductor systems, we want to emphasize the necessity of using the solvent-aware algorithm for semiconductor systems.\\
A crucial underlying factor determining SCF convergence in DFT+SCCS calculations is the solute-solvent interface calculated based on the electron density. Fig.~\ref{figs4} presents the isosurfaces of dielectric functions provided by the old CP2K implementation of the traditional Andreussi et al.'s local solute-solvent interfaces and our CP2K implementation of the non-local solvent-aware solute-solvent interfaces under conditions of significantly high $\rho^{max} = 1 \times 10^{-1}$ and $\rho^{min} = 1 \times 10^{-2}$, calculated from the electron density during the third step of SCF at which SCCS was still not activated. When SCF fails to converge, isolated solvent cavity regions within the solute are often observed. Therefore, in situations where SCF does not converge, the first step should be to investigate the rationality of the solute-solvent interface, particularly the spatial distribution of dielectric functions. Additionally, while extensive invasion of implicit solvent into the solute is unreasonable, it does not necessarily preclude SCF convergence, as demonstrated in the cases of liquid H\(_2\)O bulk and surface (Figs.~\ref{fig1a} to~\ref{fig1d}) and the (110) rutile TiO\textsubscript{2} surface (Fig.~\ref{figs4j}). Small, cavity-like implicit solvent regions that occasionally arise and cannot be properly eliminated during the optimization process are more likely to cause SCF divergence. Furthermore, although incorrect implicit solvent regions may not be effectively eliminated (and sometimes may not be altered at all by the solvent-aware algorithm as shown in Figs.~\ref{figs4d} to~\ref{figs4h}) when using large $\rho^{max}$ and $\rho^{min}$ parameters, the effectiveness of the solvent-aware algorithm heavily depends on the selected $R_{solv}$ and $f_0$ parameters. Our test results suggest that by appropriately adjusting these parameters, incorrect implicit solvent regions within the solute can often be well eliminated.\\
Tables~\ref{figs5} and~\ref{figs6} show the results of the aforementioned tests on SCF convergence using the Broyden density mixing approach and the OT SCF optimization approach, respectively, instead of the modified Pulay mixing approach. Both tables clearly demonstrate that SCF convergence worsens with an increase in the $\rho^{max}$ and $\rho^{min}$ parameters. Additionally, general improvements in SCF convergence were observed when using the solvent-aware interface compared to the traditional Andreussi et al.'s local solute-solvent interface. By comparing Tables~\ref{fig4} and~\ref{figs5}, it is evident that the SCF convergence achieved with the modified Pulay density mixing method is systematically better than that provided by the Broyden method across various systems and combinations of $\rho^{max}$ and $\rho^{min}$ parameters. As shown in Table \ref{figs6}, the OT method yields good SCF convergence for TiO\textsubscript{2} and H\textsubscript{2}O systems, but poorer SCF convergence for Pt surface systems.\\
\section{Conclusion}
\label{sec5}
We rigorously derived the solvent-aware algorithm, a modified SCCS approach based on a non-local solute-solvent interface recently introduced by Andreussi et al.\cite{andreussi_solvent-aware_2019}, from the most primitive starting point using the mathematical tools of functional analysis, and implemented it into the CP2K software package. Our derivation, which was not presented in any literature to the best of our knowledge, and implementation encompass both the potential and the analytical force terms compatible with the Quickstep framework of the CP2K package. This algorithm can eliminate erroneous pockets or cavity regions in the solute body, which are identified as interfacial transition or solvent regions by the Andreussi et al.'s local solute-solvent interface method. As a result, the convergence behaviors of the DFT+SCCS SCF iterations are significantly improved.\\
We performed a series of tests to validate our implementations. We validated the analytical force expression and its implementation by comparing the analytical atomic forces calculated with our implementation to the numerical atomic forces calculated using a finite difference approach. Depending on the choice of the parameters \(\rho^{min}\) and \(\rho^{max}\), the maximum difference between the analytical and numerical forces lies within the order of \(1.0 \times 10^{-7}\) or \(1.0 \times 10^{-4}\) Hartree/Bohr over the range of tested displacements. To evaluate the outcomes of our modified SCCS implementation, which is based on the non-local solvent-aware solute-solvent interface, we calculated the dielectric functions and compared these with the ones predicted by the older implementation, which uses Andreussi et al.'s local solute-solvent interface, across a variety of systems using the same electron densities. The test results show that our implementation can successfully eliminate implicit solvent regions that invade the pure solute region. Furthermore, we tested the impact of the two core parameters, $f_{0}$ and $R_{solv}$, on the solute-solvent interface predicted by the solvent-aware algorithm. The test results are consistent with the analysis of the effects of parameter variations on the solute-solvent interface. The tests also demonstrate that by adjusting these two parameters, our implementation can effectively modulate the degree of correction to the solute-solvent interface predicted by the traditional Andreussi et al.'s local solute-solvent interface method. Lastly, we tested and valided the improvement in SCF convergence achieved by the solvent-aware algorithm over the older SCCS implementation based on the traditional Andreussi et al.'s local solute-solvent interface. Additionally, the test results indicate that the density mixing method or different SCF optimization schemes also significantly impact the convergence of DFT+SCCS.\\

\section*{Acknowledgement}
This work was supported by the University of Zurich and SNSF Sinergia Project CRSII5\_202225. This work was supported by a grant from the Swiss National Supercomputing Centre (CSCS) under project ID s1216. We thank Dr. Johann Mattiat for discussing the acceleration of convolution calculations using FFT in CP2K with us. We thank Dr. Luis Ignacio Hernandez Segura for helping with checking formulas.

\appendix
\section{}
\label{app}
\setcounter{equation}{0}
\renewcommand\theequation{S\arabic{equation}}
\setcounter{figure}{0}
\renewcommand\thefigure{S\arabic{figure}}
\setcounter{table}{0}
\renewcommand\thetable{S\arabic{table}}
\subsection{The derivation of the equivalence of the second and third terms in Eq. \ref{eq26}}
\label{app1}
According to the product rule: $\nabla \cdot(\varphi \boldsymbol{F})=\nabla \varphi \cdot \boldsymbol{F}+\varphi \nabla \cdot \boldsymbol{F}$ (in which $\nabla \cdot$ is the divergence operator $\left(\nabla \cdot \boldsymbol{F}=\frac{\partial F_x}{\partial x}+\frac{\partial F_y}{\partial y}+\frac{\partial F_z}{\partial z}\right), \nabla$ is the gradient operator $\left(\nabla \varphi=\frac{\partial \varphi_x}{\partial x} \overrightarrow{\boldsymbol{e}_{\boldsymbol{x}}}+\frac{\partial \varphi_y}{\partial y} \overrightarrow{\boldsymbol{e}_{\boldsymbol{y}}}+\frac{\partial \varphi_z}{\partial z} \overrightarrow{\boldsymbol{e}_{\boldsymbol{z}}}\right), \varphi$ is a scalar-valued function, and $\boldsymbol{F}$ is a vector field,) one can derive:
\begin{equation}\label{eqs16}
\begin{aligned}
& \frac{1}{8 \pi} \int \varepsilon(\boldsymbol{r})\left|\nabla \phi^{t o t}(\boldsymbol{r})\right|^2 d \boldsymbol{r} \\
& \quad=\frac{1}{8 \pi} \int \nabla \cdot\left(\epsilon(\boldsymbol{r}) \nabla \phi^{t o t}(\boldsymbol{r}) \phi^{t o t}(\boldsymbol{r})\right) d \boldsymbol{r}-\frac{1}{8 \pi} \int \nabla \cdot\left(\epsilon(\boldsymbol{r}) \nabla \phi^{t o t}(\boldsymbol{r})\right) \phi^{t o t}(\boldsymbol{r}) d \boldsymbol{r} .
\end{aligned}
\end{equation}
Using divergence theorem and neglecting the surface term, one can have:
\begin{equation}\label{eqs17}
\frac{1}{8 \pi} \int \varepsilon(\boldsymbol{r})\left|\nabla \phi^{t o t}(\boldsymbol{r})\right|^2 d \boldsymbol{r}=-\frac{1}{8 \pi} \int \nabla \cdot\left(\epsilon(\boldsymbol{r}) \nabla \phi^{t o t}(\boldsymbol{r})\right) \phi^{t o t}(\boldsymbol{r}) d \boldsymbol{r}.
\end{equation}
According to the generalized Poisson equation in Eq. \ref{eq20}, $\frac{1}{8 \pi} \int \varepsilon(\boldsymbol{r})\left|\nabla \phi^{t o t}(\boldsymbol{r})\right|^2 d \boldsymbol{r}$ can be rewritten as:
\begin{equation}\label{eqs18}
\frac{1}{8 \pi} \int \varepsilon(\boldsymbol{r})\left|\nabla \phi^{t o t}(\boldsymbol{r})\right|^2 d \boldsymbol{r}=\frac{1}{2} \int \rho^{solute}(\boldsymbol{r}) \phi^{tot}(\boldsymbol{r}) d \boldsymbol{r}.
\end{equation}
\subsection{The derivation of the analytical force in the case of Andreussi et al.'s local solute-solvent interface}
\label{app2}
In the case that the local $\epsilon(\boldsymbol{r})$ in Eqs. \ref{eqs1}--\ref{eqs3} is adopted, the analytical force expression given in Eq. \ref{eqs33} becomes:
\begin{equation}\label{eqs36}
f_i^A=\frac{1}{8 \pi} \int \frac{\partial \epsilon(\boldsymbol{r})}{\partial R_i^A}\left|\nabla \phi^{tot}(\boldsymbol{r})\right|^2 d \boldsymbol{r}-\int \frac{\partial \rho^{solute}(\boldsymbol{r})}{\partial R_i^A} \phi^{tot}(\boldsymbol{r}) d \boldsymbol{r} .
\end{equation}
Since $\epsilon(\boldsymbol{r})$ is a function of $\rho^{el}(\boldsymbol{r})$, Eq. \ref{eqs36} can be further derived as:
\begin{equation}\label{eqs37}
\begin{aligned}
& f_i^A=\frac{1}{8 \pi} \int \frac{\partial \epsilon(\boldsymbol{r})}{\partial R_i^A}\left|\nabla \phi^{t o t}(\boldsymbol{r})\right|^2 d \boldsymbol{r}-\int \frac{\partial \rho^{solute}(\boldsymbol{r})}{\partial R_i^A} \phi^{t o t}(\boldsymbol{r}) d \boldsymbol{r} \\
& =\frac{1}{8 \pi} \int \frac{\partial \epsilon}{\partial \rho^{e l}}(\boldsymbol{r}) \frac{\partial \rho^{e l}(\boldsymbol{r})}{\partial R_i^A}\left|\nabla \phi^{t o t}(\boldsymbol{r})\right|^2 d \boldsymbol{r}-\int \frac{\partial \rho^{el}(\boldsymbol{r})}{\partial R_i^A} \phi^{t o t}(\boldsymbol{r}) d \boldsymbol{r} \\
& -\int \frac{\partial n_c^A(\boldsymbol{r})}{\partial R_i^A} \phi^{t o t}(\boldsymbol{r}) d \boldsymbol{r} \\
& =-\int\left(-\frac{1}{8 \pi} \frac{\partial \epsilon}{\partial \rho^{e l}}(\boldsymbol{r})\left|\nabla \phi^{t o t}(\boldsymbol{r})\right|^2+\phi^{t o t}(\boldsymbol{r})\right) \frac{\partial \rho^{e l}(\boldsymbol{r})}{\partial R_i^A} d \boldsymbol{r} \\
& -\int \frac{\partial n_c^A(\boldsymbol{r})}{\partial R_i^A} \phi^{t o t}(\boldsymbol{r}) d \boldsymbol{r} \\
& =-\int V_{Hartree}^{new, local}(\boldsymbol{r}) \frac{\partial \rho^{el}(\boldsymbol{r})}{\partial R_i^A} d \boldsymbol{r}-\int \frac{\partial n_c^A(\boldsymbol{r})}{\partial R_i^A} \phi^{tot}(\boldsymbol{r}) d \boldsymbol{r}, \\
&
\end{aligned}
\end{equation}
in which $V_{Hartree}^{new, local}(\boldsymbol{r})$ is the contribution to the Kohn-Sham potential for the implicit solvation Hartree energy with an Andreussi et al.'s local solute-solvent interface.  
\subsection{The analytical atomic force and the finite difference force at each displacement}
\label{app3}
\begin{figure}[H]
\centering
\subfigure[]{
\includegraphics[width=0.45\linewidth]{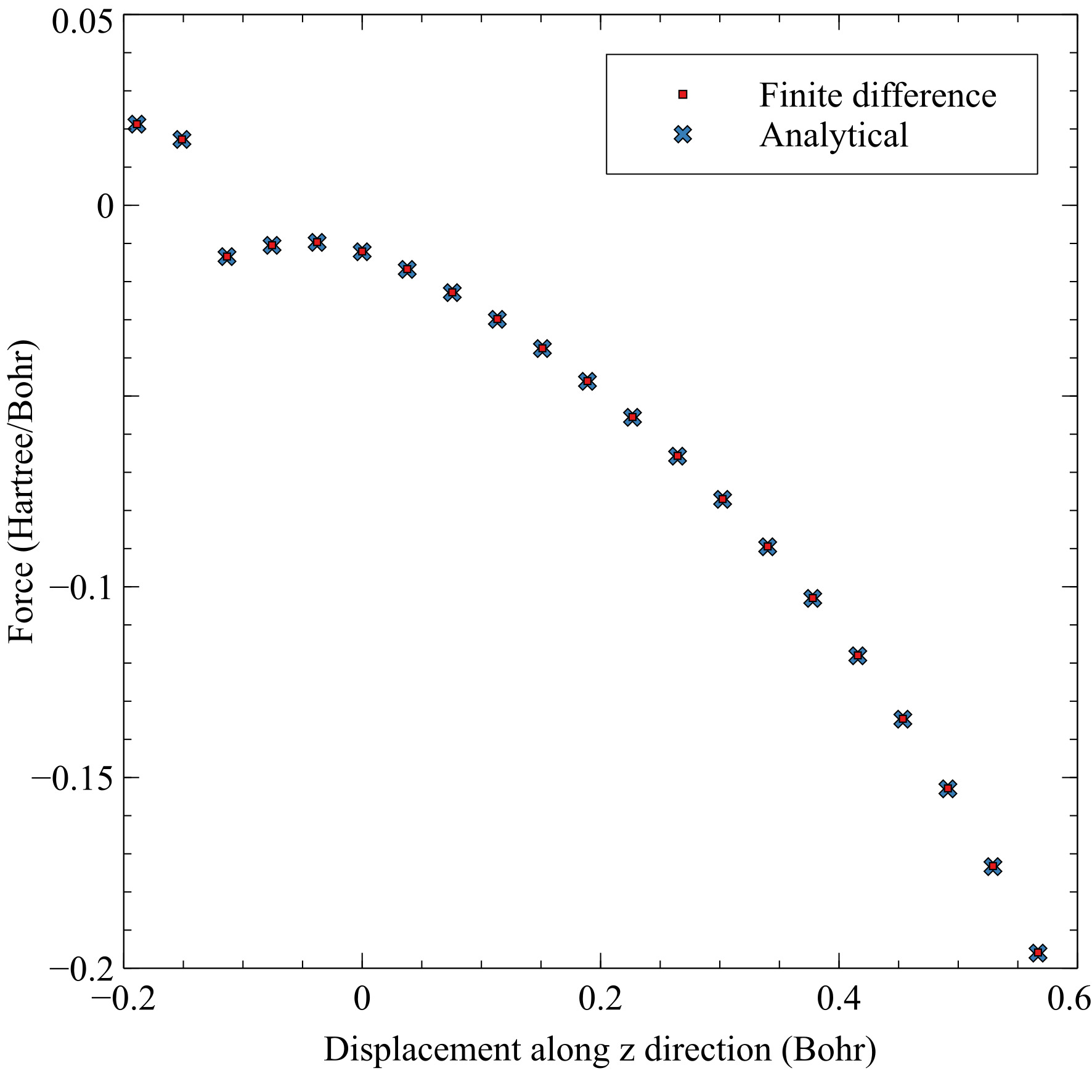}
\label{figs3a}
}
\quad
\subfigure[]{
\includegraphics[width=0.45\linewidth]{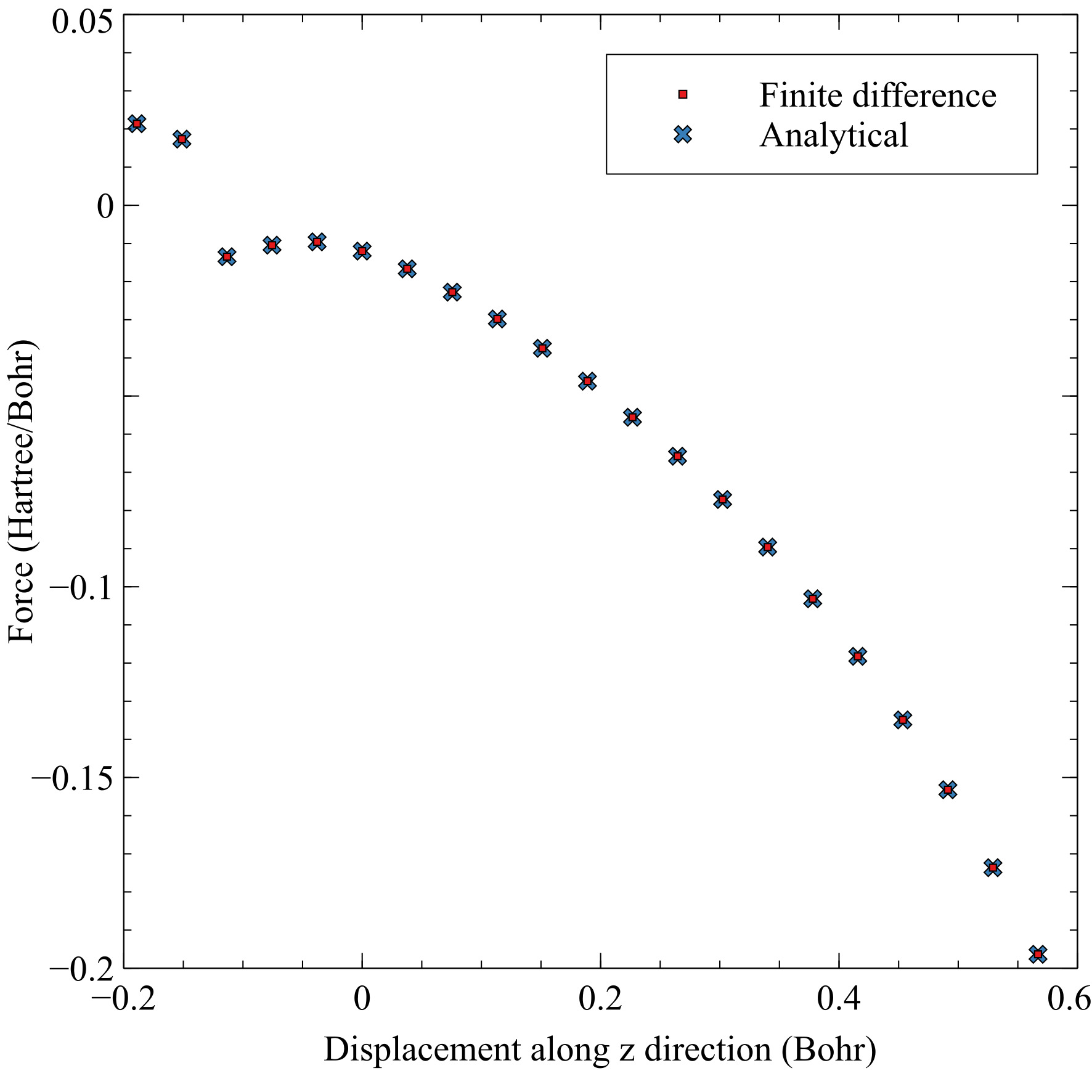}
\label{figs3b}
}
\quad
\subfigure[]{
\includegraphics[width=0.45\linewidth]{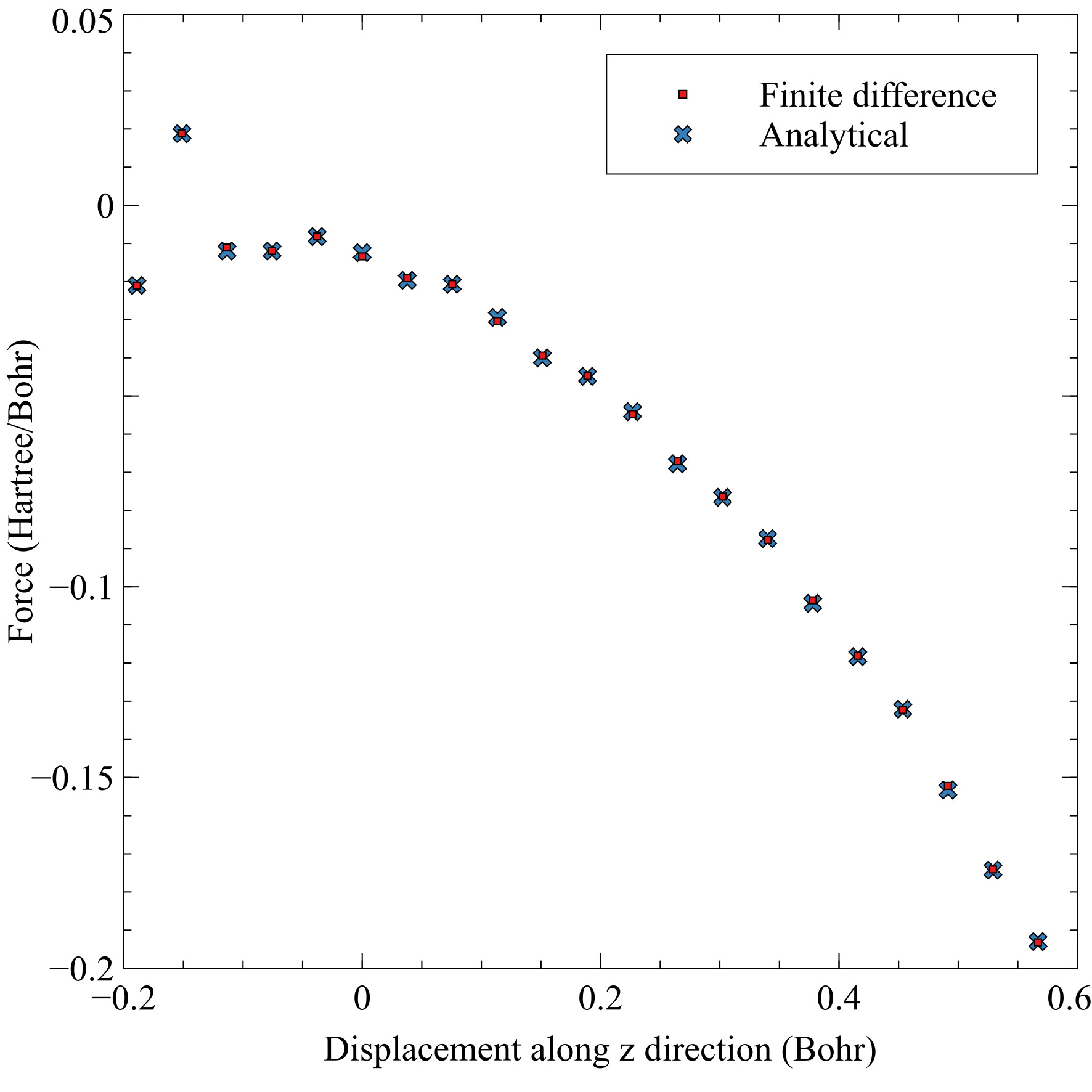}
\label{figs3c}
}
\quad
\caption{Total atomic forces on the Pt atom in the outermost layer of the Pt (111) surface when it was displaced along $\vec{r}=(1,1,1)$ direction. Small square signs represent finite difference forces. The cross signs represent the analytical forces. (a) DFT+SCCS based on the solvent-aware algorithm with the solute-solvent boundary parameters: {$\rho^{max }=1.32 \times 10^{-2}$, $\rho^{min }=5.1 \times 10^{-4}$ }. (b) DFT+SCCS based on the local Andreussi et al.'s solute-solvent boundary (Eqs. \ref{eqs1}--\ref{eqs3} in this paper) with the parameters: {$\rho^{max }=1 \times 10^{-3}$, $\rho^{min }=1 \times 10^{-4}$ }. (c) DFT+SCCS based on the local Andreussi et al.'s solute-solvent boundary (Eqs. \ref{eqs1}--\ref{eqs3} in this paper) with the parameters: {$\rho^{max }=1.32 \times 10^{-2}$, $\rho^{min }=5.1 \times 10^{-4}$ }}\label{figs3}
\end{figure}
\subsection{The differences between the analytical forces and the corresponding finite difference forces}
\label{app4}
\begin{table}[H]
\centering
\caption{DFT+SCCS based on the solvent-aware algorithm with the solute-solvent boundary parameters: {$\rho^{max }=1 \times 10^{-3}$, $\rho^{min }=1 \times 10^{-4}$ }.}\label{tabs1a}
\begin{tabular}{|c|c|}
\hline Displacement (Bohr) & Force difference (Hartree/Bohr)\\
\hline-0.1 & $-6.3 \times 10^{-8}$\\
\hline-0.08 & $-9.5 \times 10^{-9}$\\
\hline-0.06 & $-6.9 \times 10^{-8}$\\
\hline-0.04 & $-8.3 \times 10^{-8}$\\
\hline-0.02 & $7.2 \times 10^{-8}$\\
\hline 0 & $6.3 \times 10^{-8}$\\
\hline 0.02 & $5.4 \times 10^{-9}$\\
\hline 0.04 & $-3.3 \times 10^{-8}$\\
\hline 0.06 & $-1.1 \times 10^{-8}$\\
\hline 0.08 & $-3.9 \times 10^{-8}$\\
\hline 0.1 & $1.7 \times 10^{-7}$\\
\hline 0.12 & $-2.5 \times 10^{-7}$\\
\hline 0.14 & $4.8 \times 10^{-8}$\\
\hline 0.16 & $-1.0 \times 10^{-7}$\\
\hline 0.18 & $1.4 \times 10^{-8}$\\
\hline 0.2 & $-1.1 \times 10^{-7}$\\
\hline 0.22 & $-3.1 \times 10^{-7}$\\
\hline 0.24 & $1.9 \times 10^{-9}$\\
\hline 0.26 & $-2.7 \times 10^{-7}$\\
\hline 0.28 & $-1.8 \times 10^{-7}$\\
\hline 0.3 & $-3.3 \times 10^{-7}$\\
\hline
\end{tabular}
\end{table}
\begin{table}[H]
\centering
\caption{DFT+SCCS based on the solvent-aware algorithm with the solute-solvent boundary parameters: {$\rho^{max }=1.32 \times 10^{-2}$, $\rho^{min }=5.1 \times 10^{-4}$ }.}\label{tabs1b}
\begin{tabular}{|c|c|}
\hline Displacement (Bohr) & $\mid$Force difference$\mid$ (Hartree/Bohr) \\
\hline-0.1 & $-5.5 \times 10^{-6}$\\
\hline-0.08 & $-1.0 \times 10^{-5}$\\
\hline-0.06 & $-1.9 \times 10^{-5}$\\
\hline-0.04 & $-4.4 \times 10^{-5}$\\
\hline-0.02 & $-7.2 \times 10^{-5}$\\
\hline 0 & $-8.2 \times 10^{-5}$\\
\hline 0.02 & $-7.8 \times 10^{-5}$\\
\hline 0.04 & $-7.5 \times 10^{-5}$\\
\hline 0.06 & $-7.2 \times 10^{-5}$\\
\hline 0.08 & $-7.3 \times 10^{-5}$\\
\hline 0.1 & $-7.3 \times 10^{-5}$\\
\hline 0.12 & $-7.5 \times 10^{-5}$\\
\hline 0.14 & $-7.8 \times 10^{-5}$\\
\hline 0.16 & $-8.1 \times 10^{-5}$\\
\hline 0.18 & $-8.6 \times 10^{-5}$\\
\hline 0.2 & $-9.2 \times 10^{-5}$\\
\hline 0.22 & $-1.0 \times 10^{-4}$\\
\hline 0.24 & $-1.1 \times 10^{-4}$\\
\hline 0.26 & $-1.2 \times 10^{-4}$\\
\hline 0.28 & $-1.4 \times 10^{-4}$\\
\hline 0.3 & $-1.5 \times 10^{-4}$\\
\hline
\end{tabular}
\end{table}
\begin{table}[H]
\centering
\caption{DFT+SCCS based on the local Andreussi et al.'s solute-solvent boundary (Eqs. \ref{eqs1}--\ref{eqs3} in this paper) with the parameters: {$\rho^{max }=1 \times 10^{-3}$, $\rho^{min }=1 \times 10^{-4}$ }.}\label{tabs1c}
\begin{tabular}{|c|c|}
\hline Displacement (Bohr) & $\mid$Force difference$\mid$ (Hartree/Bohr) \\
\hline-0.1 & $-3.3 \times 10^{-8}$\\
\hline-0.08 & $1.9 \times 10^{-8}$\\
\hline-0.06 & $2.6 \times 10^{-8}$\\
\hline-0.04 & $-4.9 \times 10^{-8}$\\
\hline-0.02 & $1.5 \times 10^{-7}$\\
\hline 0 & $1.0 \times 10^{-6}$\\
\hline 0.02 & $-1.7 \times 10^{-7}$\\
\hline 0.04 & $-7.4 \times 10^{-7}$\\
\hline 0.06 & $-6.3 \times 10^{-7}$\\
\hline 0.08 & $1.7 \times 10^{-7}$\\
\hline 0.1 & $-2.5 \times 10^{-7}$\\
\hline 0.12 & $-6.5 \times 10^{-8}$\\
\hline 0.14 & $5.1 \times 10^{-7}$\\
\hline 0.16 & $-3.1 \times 10^{-7}$\\
\hline 0.18 & $-3.8 \times 10^{-7}$\\
\hline 0.2 & $2.3 \times 10^{-7}$\\
\hline 0.22 & $1.1 \times 10^{-7}$\\
\hline 0.24 & $-1.1 \times 10^{-7}$\\
\hline 0.26 & $4.5 \times 10^{-7}$\\
\hline 0.28 & $3.6 \times 10^{-8}$\\
\hline 0.3 & $-1.1 \times 10^{-7}$\\
\hline
\end{tabular}
\end{table}
\begin{table}[H]
\centering
\caption{DFT+SCCS based on the local Andreussi et al.'s solute-solvent boundary (Eqs. \ref{eqs1}--\ref{eqs3} in this paper) with the parameters: {$\rho^{max }=1.32 \times 10^{-2}$, $\rho^{min }=5.1 \times 10^{-4}$ }}\label{tabs1d}
\begin{tabular}{|c|c|}
\hline Displacement (Bohr) & $\mid$Force difference$\mid$ (Hartree/Bohr) \\
\hline-0.1 & $3.1 \times 10^{-5}$\\
\hline-0.08 & $-4.0 \times 10^{-5}$\\
\hline-0.06 & $-9.5 \times 10^{-4}$\\
\hline-0.04 & $-6.9 \times 10^{-5}$\\
\hline-0.02 & $-6.0 \times 10^{-5}$\\
\hline 0 & $9.8 \times 10^{-4}$\\
\hline 0.02 & $-5.4 \times 10^{-4}$\\
\hline 0.04 & $-6.0 \times 10^{-5}$\\
\hline 0.06 & $9.6 \times 10^{-4}$\\
\hline 0.08 & $-6.0 \times 10^{-4}$\\
\hline 0.1 & $-1.1 \times 10^{-4}$\\
\hline 0.12 & $6.3 \times 10^{-4}$\\
\hline 0.14 & $-6.6 \times 10^{-4}$\\
\hline 0.16 & $-1.9 \times 10^{-4}$\\
\hline 0.18 & $-3.5 \times 10^{-4}$\\
\hline 0.2 & $-8.1 \times 10^{-4}$\\
\hline 0.22 & $-2.1 \times 10^{-4}$\\
\hline 0.24 & $2.0 \times 10^{-4}$\\
\hline 0.26 & $-1.1 \times 10^{-3}$\\
\hline 0.28 & $-1.6 \times 10^{-4}$\\
\hline 0.3 & $1.7 \times 10^{-4}$\\
\hline
\end{tabular}
\end{table}

\subsection{}
\label{app5}
\begin{figure}[H]
\centering
\subfigure[]{
\includegraphics[height=0.13\textwidth]{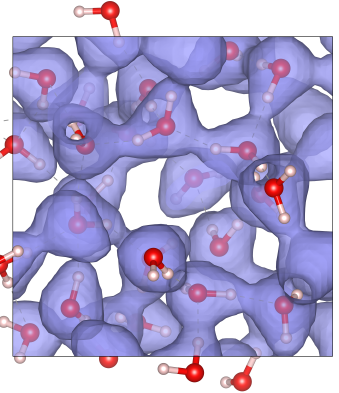}
\label{figs4a}
}
\hspace{0pt}
\subfigure[]{
\includegraphics[height=0.13\textwidth]{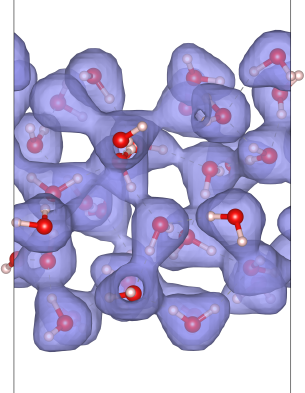}
\label{figs4c}
}
\hspace{0pt}
\subfigure[]{
\includegraphics[height=0.13\textwidth]{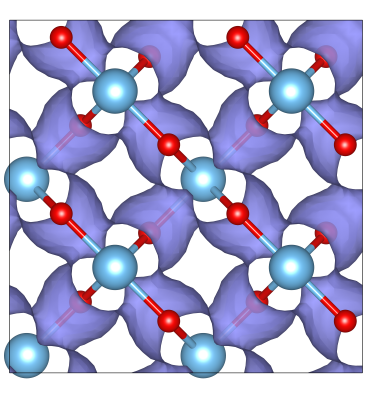}
\label{figs4e}
}
\hspace{0pt}
\subfigure[]{
\includegraphics[height=0.13\textwidth]{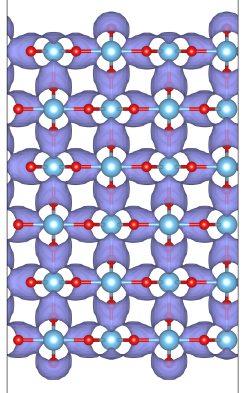}
\label{figs4g}
}
\hspace{0pt}
\subfigure[]{
\includegraphics[height=0.13\textwidth]{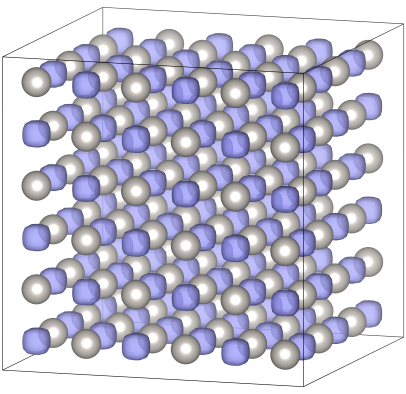}
\label{figs4i}
}
\hspace{0pt}
\subfigure[]{
\includegraphics[height=0.13\textwidth]{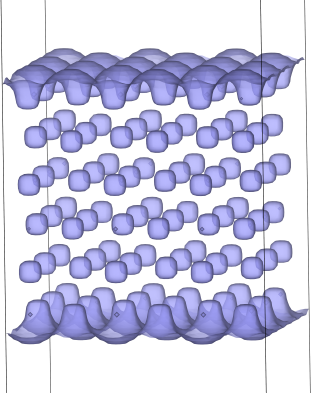}
\label{figs4k}
}
\hspace{0pt}
\subfigure[]{
\includegraphics[height=0.13\textwidth]{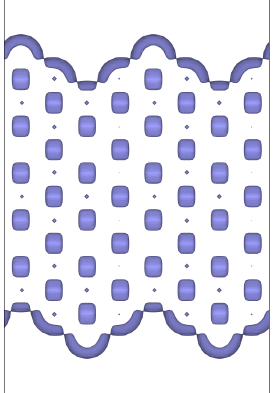}
\label{figs4m}
}
\hspace{0pt}
\subfigure[]{
\includegraphics[height=0.13\textwidth]{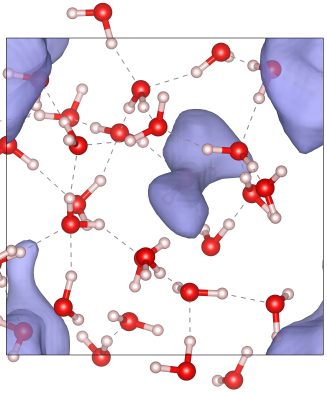}
\label{figs4b}
}
\hspace{0pt}
\subfigure[]{
\includegraphics[height=0.13\textwidth]{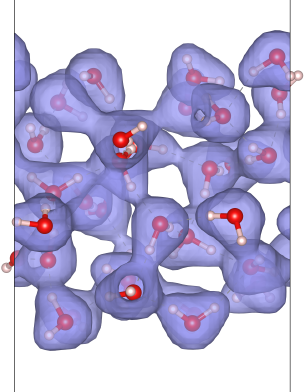}
\label{figs4d}
}
\hspace{0pt}
\subfigure[]{
\includegraphics[height=0.13\textwidth]{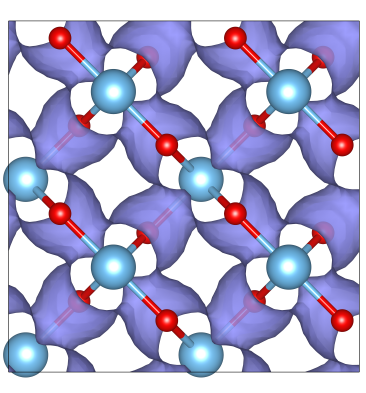}
\label{figs4f}
}
\hspace{0pt}
\subfigure[]{
\includegraphics[height=0.13\textwidth]{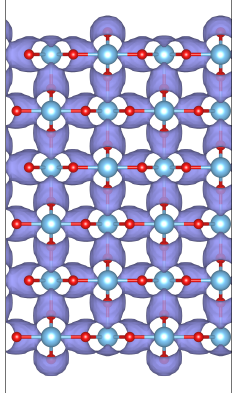}
\label{figs4h}
}
\hspace{0pt}
\subfigure[]{
\includegraphics[height=0.13\textwidth]{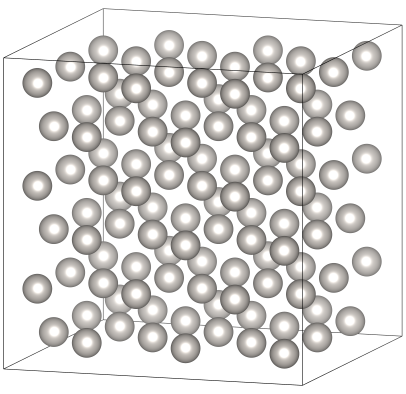}
\label{figs4j}
}
\hspace{0pt}
\subfigure[]{
\includegraphics[height=0.13\textwidth]{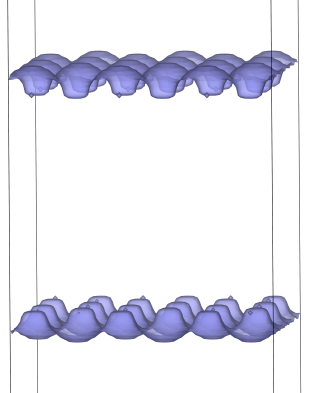}
\label{figs4l}
}
\hspace{0pt}
\subfigure[]{
\includegraphics[height=0.13\textwidth]{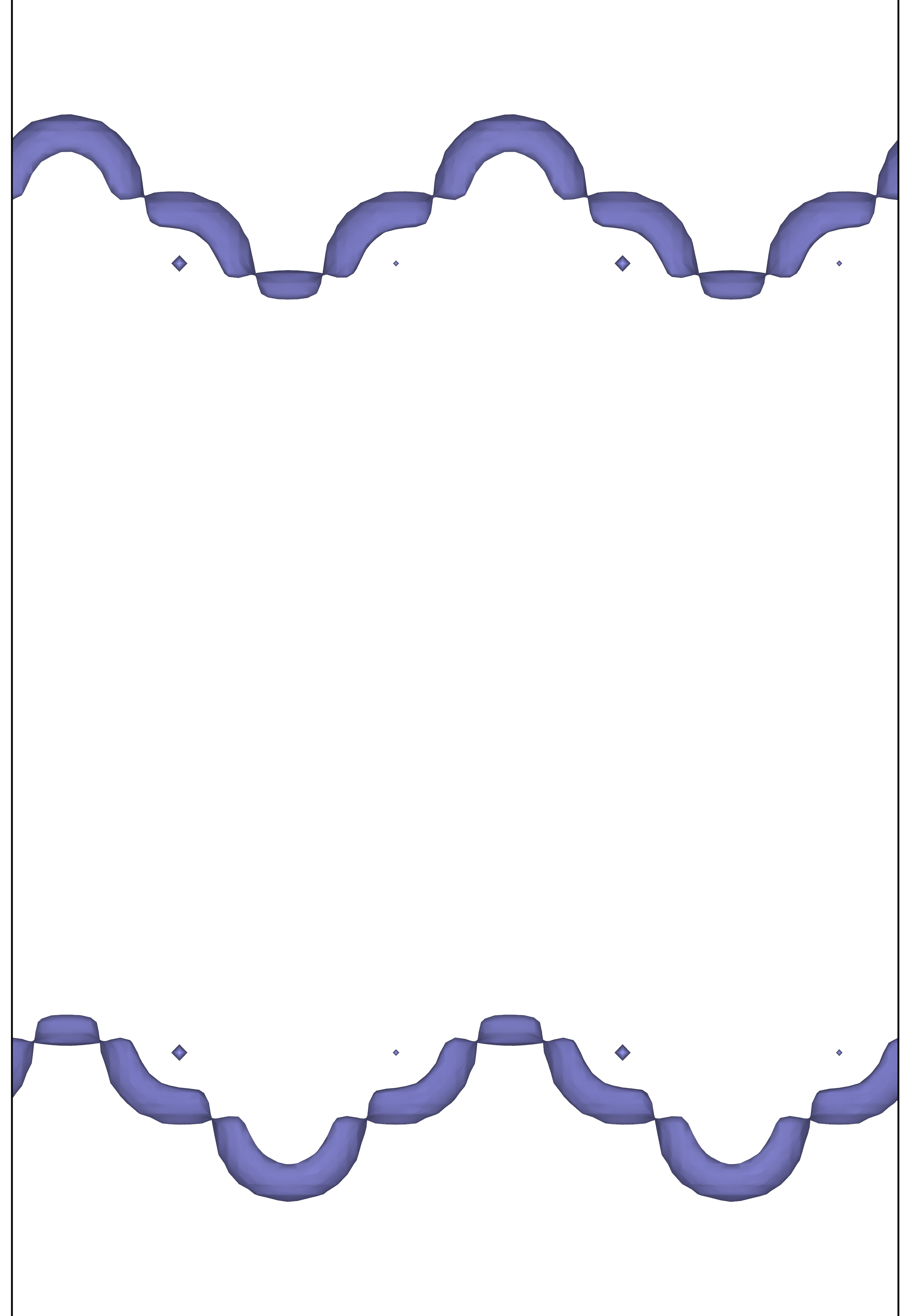}
\label{figs4n}
}
\quad
\caption{
Isosurfaces of the dielectric functions calculated using the traditional Andreussi et al.'s local solute-solvent interface algorithm or the newly implemented solvent-aware algorithm (with the isosurface level of 13), respectively. The upper half of the figure presents results from the old local interface implementation (\subref{figs4a}--\subref{figs4m}), while the lower half shows results from the implementation of the non-local solvent-aware algorithm (\subref{figs4b}--\subref{figs4n}). From top to bottom, each pair of sub-figures corresponds to the same test system: the liquid H\(_2\)O bulk (\subref{figs4a} and \subref{figs4b}), the liquid H\(_2\)O surface (\subref{figs4c} and \subref{figs4d}), the rutile TiO\textsubscript{2} bulk (\subref{figs4e} and \subref{figs4f}), the (110) rutile TiO\textsubscript{2} surface (\subref{figs4g} and \subref{figs4h}), the Pt bulk (\subref{figs4i} and \subref{figs4j}), the (100) surface for Pt (\subref{figs4k} and \subref{figs4l}), and the “missing row” reconstructed (110) surface for Pt (\subref{figs4m} and \subref{figs4n}), all calculated using $\rho^{max }=1 \times 10^{-1}$, $\rho^{min }=1 \times 10^{-2}$ density threshold values. In the case of the Pt surface systems, the atomistic model of the surfaces has been hidden to facilitate better visual observation of the isosurfaces.}\label{figs4}
\end{figure}

\subsection{}
\label{app6}
\begin{figure}[H]
\centering
\includegraphics[width=\textwidth]{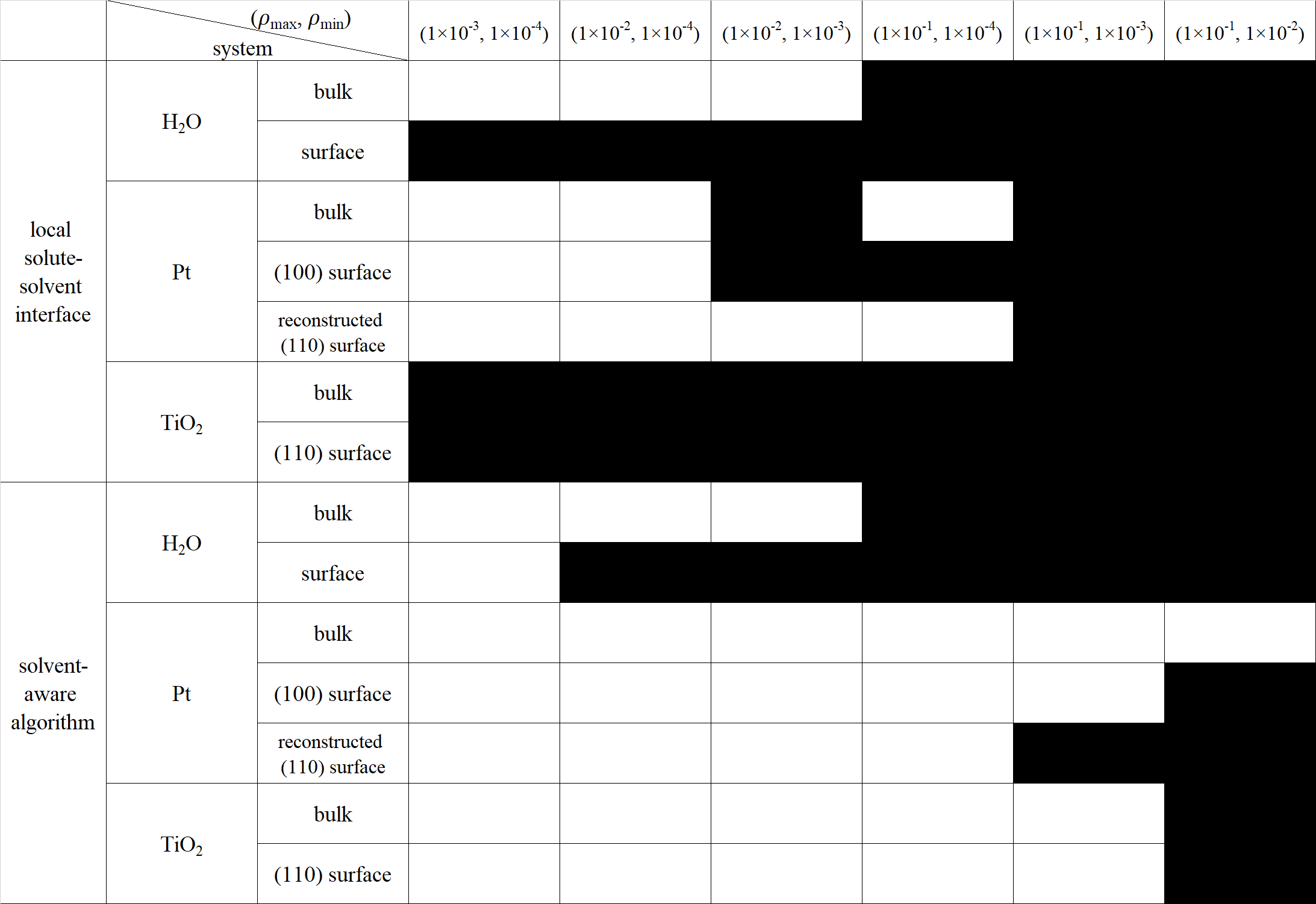}
\quad
\captionsetup{name=Table}
\renewcommand{\thefigure}{S5}
\caption{The SCF convergence of DFT+SCCS calculations within 500 iterations under various $\rho^{max}$ and $\rho^{min}$ parameters, using both the old CP2K SCCS implementation based on the Andreussi et al.'s local solute-solvent interface approach and our CP2K implementation of the modified SCCS based on the non-local solvent-aware solute-solvent interfaces. The diagonalization approach and the Broyden mixing approach were used in all calculations.  In the table, white and black indicate the convergence and divergence of SCF iterations within 500 steps, respectively.}
\renewcommand{\thefigure}{\arabic{figure}}
\label{figs5}
\end{figure}

\begin{figure}[H]
\centering
\includegraphics[width=\textwidth]{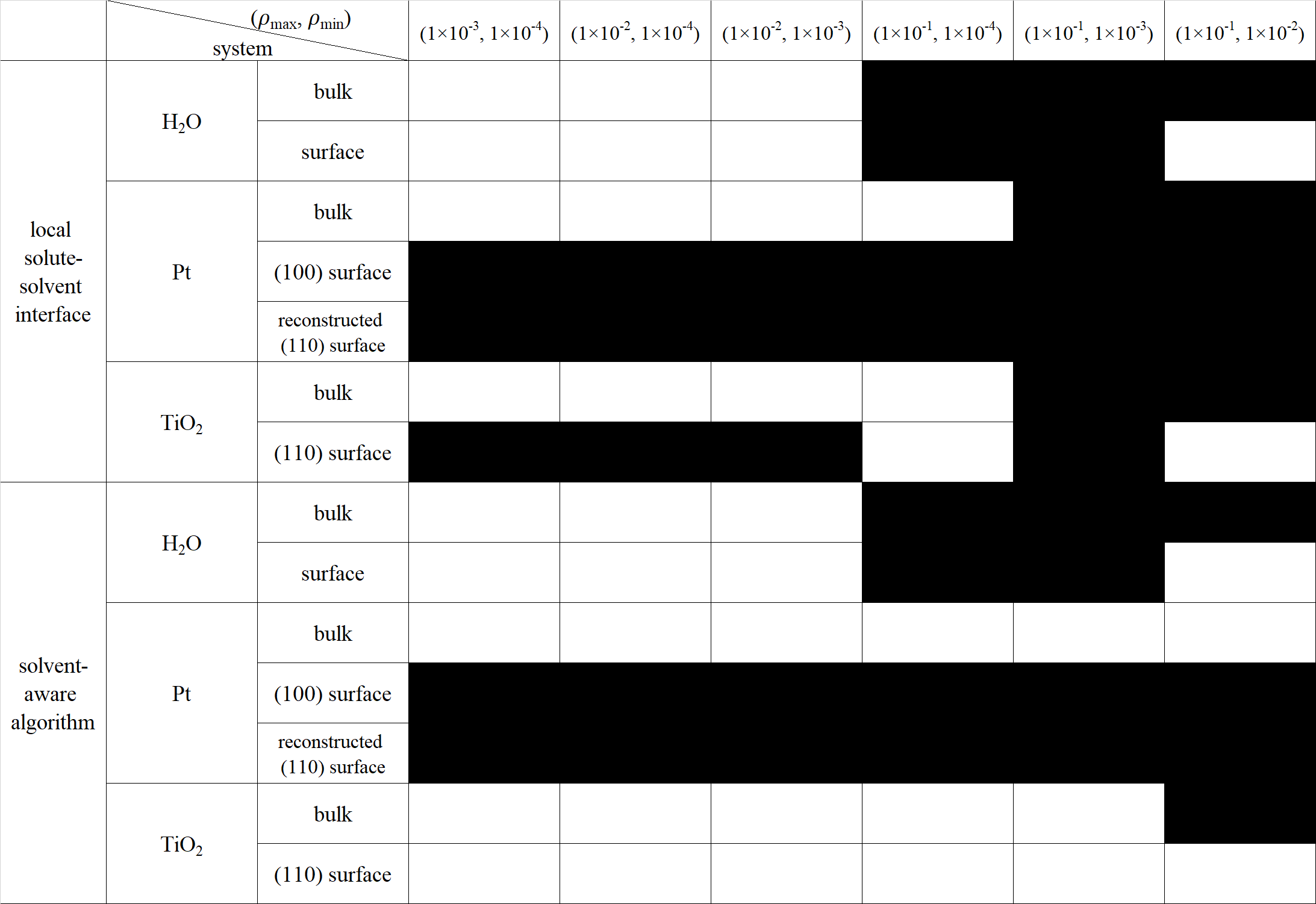}
\quad
\captionsetup{name=Table}
\renewcommand{\thefigure}{S6}
\caption{
The SCF convergence of DFT+SCCS calculations within 500 iterations under various $\rho^{max}$ and $\rho^{min}$ parameters, using both the old CP2K SCCS implementation based on the Andreussi et al.'s local solute-solvent interface approach and our CP2K implementation of the modified SCCS based on the non-local solvent-aware solute-solvent interfaces. The orbital transformation SCF optimization approach was used in all calculations. In the table, white and black indicate the convergence and divergence of SCF iterations within 500 steps, respectively.}
\renewcommand{\thefigure}{\arabic{figure}}
\label{figs6}
\end{figure}
\subsection{}
\label{app7}
The general chain rule is, $\frac{\delta i(\boldsymbol{r})_{\left[ k(\boldsymbol{r}'')_{\left[j(\boldsymbol{r}')\right]}\right]}}{\delta j(\boldsymbol{r}^{\prime})} = \int \frac{\delta i(\boldsymbol{r})_{\left[ k(\boldsymbol{r}'')_{\left[j(\boldsymbol{r}')\right]}\right]}}{\delta k(\boldsymbol{r}^{\prime\prime})_{\left[j(\boldsymbol{r}^{\prime})\right]}} \frac{\delta k(\boldsymbol{r}^{\prime\prime})_{\left[j(\boldsymbol{r}^{\prime})\right]}}{\delta j(\boldsymbol{r}^{\prime})} \, d\boldsymbol{r}^{\prime\prime}$. If $k$ is a local function of $j$, the chain rule is $\frac{\delta i(\boldsymbol{r})_{\left[k(j(\boldsymbol{r}'))\right]}}{\delta j(\boldsymbol{r}')} = \frac{\delta i(\boldsymbol{r})_{\left[k(j(\boldsymbol{r}'))\right]}}{\delta k(j(\boldsymbol{r}'))} \frac{dk(j)}{dj}(\boldsymbol{r}')$.\\

$s\left(\boldsymbol{r}^{\prime \prime}\right)$ is equal to $\int s\left(\boldsymbol{r}^{\prime \prime \prime}\right) \delta\left(\boldsymbol{r}^{\prime \prime \prime}-\boldsymbol{r}^{\prime \prime}\right) d \boldsymbol{r}^{\prime \prime \prime}$. According to the definition of a functional derivative, one can derive\\
$\int \frac{\delta s\left(\boldsymbol{r}^{\prime \prime}\right)}{\delta s\left(\boldsymbol{r}^{\prime \prime \prime}\right)} f\left(\boldsymbol{r}^{\prime \prime \prime}\right) d \boldsymbol{r}^{\prime \prime \prime}\\
=\lim _{\epsilon \rightarrow 0} \frac{\left(\int\left(s\left(\boldsymbol{r}^{\prime \prime \prime}\right)+\epsilon \cdot f\left(\boldsymbol{r}^{\prime \prime \prime}\right)\right) \delta\left(\boldsymbol{r}^{\prime \prime \prime}-\boldsymbol{r}^{\prime \prime}\right) d \boldsymbol{r}^{\prime \prime \prime}\right)-\left(\int s\left(\boldsymbol{r}^{\prime \prime \prime}\right) \delta\left(\boldsymbol{r}^{\prime \prime \prime}-\boldsymbol{r}^{\prime \prime}\right) d \boldsymbol{r}^{\prime \prime \prime}\right)}{\epsilon}\\
=\int\delta\left(\boldsymbol{r}^{\prime \prime \prime}-\boldsymbol{r}^{\prime \prime}\right) f\left(\boldsymbol{r}^{\prime \prime \prime}\right) d \boldsymbol{r}^{\prime \prime \prime}$.\\
As a result, it gives $\frac{\delta s\left(\boldsymbol{r}^{\prime \prime}\right)}{\delta s\left(\boldsymbol{r}^{\prime \prime \prime}\right)}=\delta\left(\boldsymbol{r}^{\prime \prime \prime}-\boldsymbol{r}^{\prime \prime}\right)$.\\


%%       \cite{<label>} ==> [1]

%%

\bibliographystyle{elsarticle-num} 
\bibliography{manuscript.bib}


\end{document}